\documentclass[notitlepage,nofootinbib,
twocolumn,
longbibliography,superscriptaddress]{revtex4-1}

\usepackage{amsmath}
\usepackage{amsthm, amssymb}

\usepackage{graphicx}
\usepackage[caption=false]{subfig}

\usepackage{bbold}

\usepackage[usenames,dvipsnames]{color}
\usepackage[colorlinks=true,citecolor=blue,linkcolor=magenta]{hyperref}

\usepackage{dsfont}
\usepackage{changepage}
\usepackage{array}

\newcommand{\bit}{\begin{itemize}}
\newcommand{\eit}{\end{itemize}}

\newcommand{\f}{\frac}
\renewcommand{\>}{\right\rangle}
\newcommand{\<}{\left\langle}
\newcommand{\ba}{\begin{align}}
\newcommand{\ea}{\end{align}}
\newcommand{\be}{\begin{equation}}
\newcommand{\ee}{\end{equation}}
\newcommand{\bi}{\begin{itemize}}
\newcommand{\ei}{\end{itemize}}
\newcommand{\lf}{\left(}
\newcommand{\ri}{\right)}
\newcommand{\dd}{\mathrm{d}}

\newcommand{\Tr}{\operatorname{Tr}}

\begin{document}

\newcommand{\bra}[1]{\< #1 \right|}
\newcommand{\ket}[1]{\left| #1 \>}

\title{Dynamics of entanglement and transport in 1D systems with quenched randomness}

\author{Adam Nahum}
\affiliation{Theoretical Physics, Oxford University, 1 Keble Road, Oxford OX1 3NP, United Kingdom}
\affiliation{Department of Physics, Massachusetts Institute of Technology, Cambridge, MA 02139, USA}
\author{Jonathan Ruhman}
\affiliation{Department of Physics, Massachusetts Institute of Technology, Cambridge, MA 02139, USA}
\author{David A. Huse}
\affiliation{Department of Physics, Princeton University, NJ 08544, USA}
\date{\today}

\begin{abstract}
\noindent
Quenched randomness can have a dramatic effect on the dynamics of isolated 1D quantum many-body systems, even for systems that  thermalize.  This is because transport, entanglement, and operator spreading
can be hindered by `Griffiths' rare regions which locally resemble the many-body-localized phase and thus act as weak links.  We propose coarse-grained models for entanglement growth and for the spreading of quantum operators in the presence of such weak links.  We also examine entanglement growth across a single weak link numerically.  We show that these weak links have a stronger effect on entanglement growth than previously assumed: entanglement growth is sub-ballistic whenever such weak links have a power-law probability distribution at low couplings, i.e. throughout the entire thermal Griffiths phase.  We argue that the probability distribution of the entanglement entropy across a cut can be understood from a simple picture in terms of a classical surface growth model.
Surprisingly, the four length scales associated with (i) production of entanglement, (ii) spreading of conserved quantities, (iii) spreading of operators, and (iv) the width of the `front' of a spreading operator, are characterized by dynamical exponents that in general are all distinct.  Our numerical analysis of entanglement growth between weakly coupled systems may be of independent interest.
\end{abstract}

\maketitle

\section{Introduction}
\label{introduction_sec}

A basic question about a many-body quantum system, closely related to its ability to thermalize, is how effectively quantum information spreads through it. The dynamical generation of quantum entanglement, following a quantum quench from a weakly entangled state, provides one window on information spreading \cite{CardyCalabrese2005EntanglementEvolution,DeChiaraLogGrowth,BardarsonEntanglementGrowth,KimHuse2013,EntanglementRandomUnitary}.  Unitary dynamics typically generates correlations between increasingly distant degrees of freedom as time goes on.  The resulting irreversible growth in the entanglement entropy of a subsystem reveals differences between integrable, nonintegrable, disordered and many-body localized (MBL) \cite{BardarsonEntanglementGrowth,nhreview} systems.

Complementary insight into information spreading comes from considering light-cone-like effects limiting the propagation of signals and disturbances through the system \cite{LiebRobinson}.  This leads to the question of how an initially local quantum operator spreads and becomes nonlocal under Heisenberg time evolution.  Again there are important differences between clean and disordered systems.

In translationally invariant one-dimensional (1D) systems, entanglement growth and operator spreading are both typically associated with nonzero speeds \cite{CardyCalabrese2005EntanglementEvolution,DeChiaraLogGrowth,KimHuse2013,EntanglementRandomUnitary,Roberts_Shocks}.  By contrast, in the MBL phase both entanglement growth \cite{DeChiaraLogGrowth, ZnidaricMBL2008,BardarsonEntanglementGrowth,SerbynEntanglementGrowth,PhenomenologyOfFullyMBL} and operator spreading \cite{HuangMBLOTOC,FanMBLOTOC,YuChenMBLOTOC,SwingleMBLOTOC,HeMBLOTOC,ChenMBLOTOC,SlagleMBLOTOC} are associated with length scales that grow only logarithmically in time.

This paper studies a third situation: 1D systems that are disordered, but are in the thermalizing phase. In 1D, quenched randomness can strongly affect transport and information spreading even in the thermal phase.  This is because there can exist rare regions where disorder happens to be stronger than average, and which locally resemble the MBL phase.  These `Griffiths' regions act as bottlenecks or weak links, hindering the propagation both of conserved quantities and of quantum information. The effects on transport are fairly well understood: strong enough disorder leads to subdiffusive transport \cite{VoskHuseAltmanMBL,PotterVasseurParameswaranMBL,AgarwalAnomalousDiffusion}, as observed numerically \cite{BarLevNoDiffusion,  AgarwalAnomalousDiffusion, LuitzLaflorencieAlet,  NoDiffusion1D, DiffusionHeisenbergChain}. (Certain rare-region effects have also been addressed experimentally \cite{luschen2016evidence}.) Sub-ballistic entanglement and signal propagation has  been observed numerically in the Griffiths phase \cite{LuitzLaflorencieAlet,luitz2017information}.

Here we provide  long-wavelength pictures which expose the universal physics underlying entanglement growth and operator spreading in the Griffiths phase, and yield the universal scaling exponents and scaling forms governing these processes. In contrast to both clean systems and the MBL phase, we find that the lengthscales governing entanglement growth and operator growth (as measured by the so-called out-of-time-order correlator) scale with \textit{different} powers of the time. In a certain sense, entanglement growth is parametrically slower than operator spreading in the Griffiths phase.

We provide a long-distance picture for entanglement entropy growth in terms of an effective  classical surface growth problem. The height of the growing `surface', $S(x)$, is the amount of entanglement across a cut in the 1D system at position $x$; the growth of the surface is deterministic, but it is affected by quenched randomness in the local growth rates. This picture is motivated by an analogy to a simpler quantum dynamics based on a random quantum circuit \cite{EntanglementRandomUnitary}. The  basic assumptions are also substantiated independently, including with a semi-microscopic analysis of entangling across a Griffiths region.

The surface growth picture leads to the unexpected conclusion that entanglement growth is sub-ballistic throughout the entire thermal Griffiths phase, i.e. whenever Griffiths regions locally resembling the MBL phase are possible. (This requires the absence of exact nonabelian symmetries, as these prohibit MBL \cite{ChandranKhemaniLaumannSondhi,PotterVasseurSymmetry,ProtopopovHoAbanin}.)  We give expressions for the dynamical exponent governing the growth of the von Neumann entropy and for the probability distribution of this quantity at late times.

Turning to operator growth, we propose a very simple `hydrodynamic' picture for the out-of-time-order correlator in the Griffiths phase, making use of the random circuit results in Ref.~\cite{OperatorGrowth} (see also \cite{CurtPaper}). We obtain a picture involving two separate diverging length scales at long times. In fact we argue that in the Griffiths phase there are in general at least four separate dynamical exponents $z$, characterizing four length scales that can grow with time with different exponents: these are associated with entanglement growth ($z_S$), with the spreading of an operator ($z_O$),  with the width of the `front' of the OTOC ($z_W$), and with spreading of conserved quantities ($z_C$). (See the table in Sec.~\ref{sec:exponent_summary}.)  In particular, ballistic spreading of operators does not imply ballistic spreading of entanglement, contrary to previous suggestions.

Studying the thermal Griffiths phase leads naturally to questions about how two weakly-coupled systems exchange quantum information.  These questions are of general interest, outside the context of  disordered systems.  How do two weakly coupled quantum systems become entangled over time?  How do operators localized in one of the systems spread across the weak link?  We provide numerical and analytical results for generation of entanglement across the weak link, showing that the entanglement entropy is governed by a very simple scaling function and that the `weak link' can be characterized by a well-defined entanglement growth rate.  (We investigate numerically how this growth rate depends on the strength of the weak coupling and on the index $n$ of the Renyi entropy $S_n$.)  We propose a simple scaling form for the out-of-time-order correlator of an operator which spreads across a weak link between two semi-infinite 1D chains.

\tableofcontents

\section{Entanglement growth}
\label{entanglement_growth_sec}

In this section we study entanglement growth starting from a weakly entangled state: this could be the ground state of the pre-quench Hamiltonian in a quantum quench.
The specific 1D system does not matter at this stage, but an archetypal example is the Heisenberg spin chain with random couplings and fields,
 \be
 H = \sum_i J_i \vec S_i \cdot \vec S_{i+1} + \sum_i h_i S^z_i.
 \ee
(We could also consider a Floquet spin chain with effective discrete time dynamics.)  Such chains will thermalize if the randomness is not too strong.  However, thermalization may be slow as a result of `weak links'.  These may be simply weak couplings where a single bond $J_i$ is very weak.  More robustly, the weak links may be due to extended Griffiths regions where the disorder happens to be more severe, so that MBL physics arises locally \cite{AgarwalAnomalousDiffusion}.

Moving to a coarse-grained description, let us label `weak links' by an index $i$, with a given weak link located at a position $x_i$. A key assumption is that each weak link has a well-defined local \textit{entanglement rate} $\Gamma_i$,  which is the rate at which entanglement is generated across the weak link in the absence of other weak links. We will give arguments supporting this assumption in Secs.~\ref{explicit_griffiths_argument},~\ref{weak_link_sec}. The local entanglement rates are assumed to be independent random variables with a power-law distribution at small $\Gamma$:
\ba\label{prob_gamma}
P(\Gamma)  &\sim A \, \Gamma^a, & -1< a& <\infty.
\end{align}
The exponent $a$ depends on the strength of disorder and tends to $-1$ as the many-body localization transition is approached. The power-law form arises because the probability of a Griffiths region of length $\ell$, and the entangling rate associated with such a region, both decrease exponentially with $\ell$ (Sec.~\ref{explicit_griffiths_argument}).  Or the power law might arise directly from the `bare' probability distribution of the couplings $J_i$.   As usual, we will neglect the weak link's nonzero spatial extent $\ell$: for Griffiths regions this grows only logarithmically as $\Gamma\rightarrow 0$, so is negligible compared to the lengthscales discussed below which grow as powers of $\Gamma$.

\begin{figure}[t]
 \begin{center}
  \includegraphics[width=\linewidth]{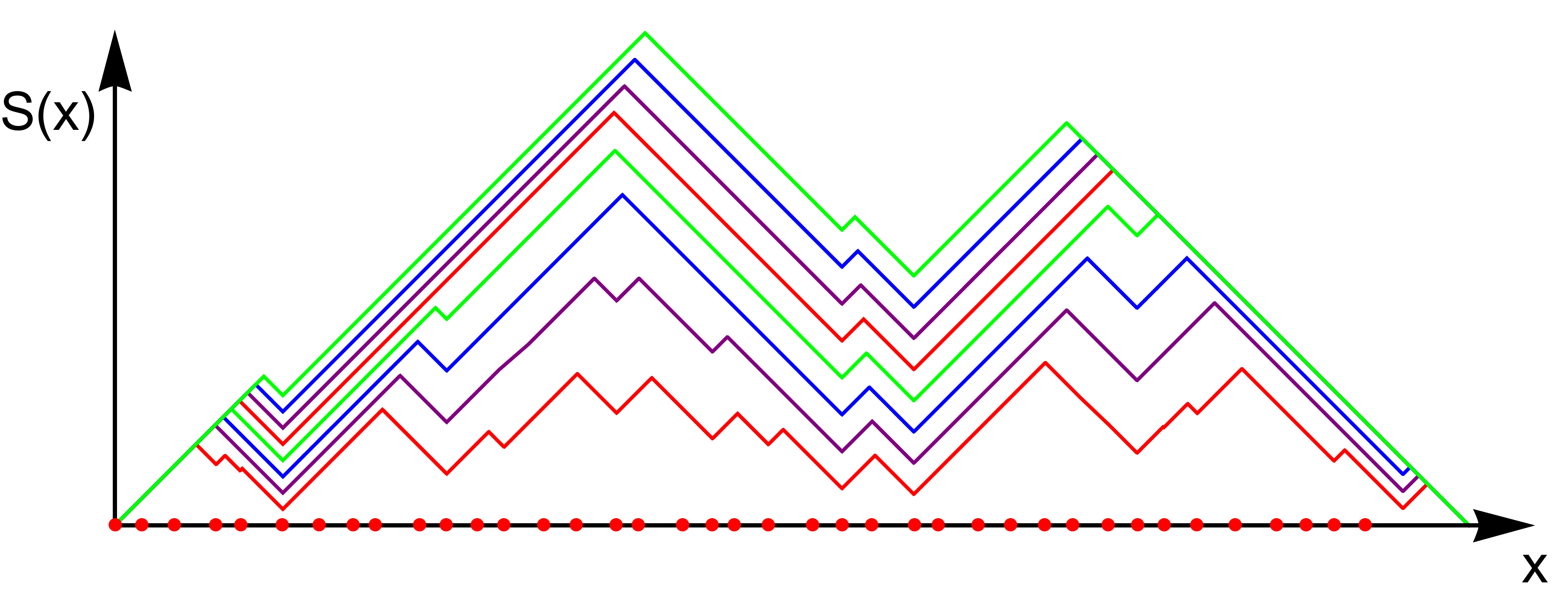}
 \end{center}
\caption{Surface growth picture for entanglement $S(x)$ across a cut at $x$ in a finite chain, following Eq.~\ref{coarse-grained growth rule}. The lines show successive equally spaced times. Weak links with larger rates $\Gamma$ are successively `dominated' by weaker links with smaller $\Gamma$. (This sample was generated with $a=0$, see Eq.~\ref{prob_gamma}.)}
 \label{growth_schematic}
\end{figure}

\begin{figure}[b]
 \begin{center}
  \includegraphics[width=0.7\linewidth]{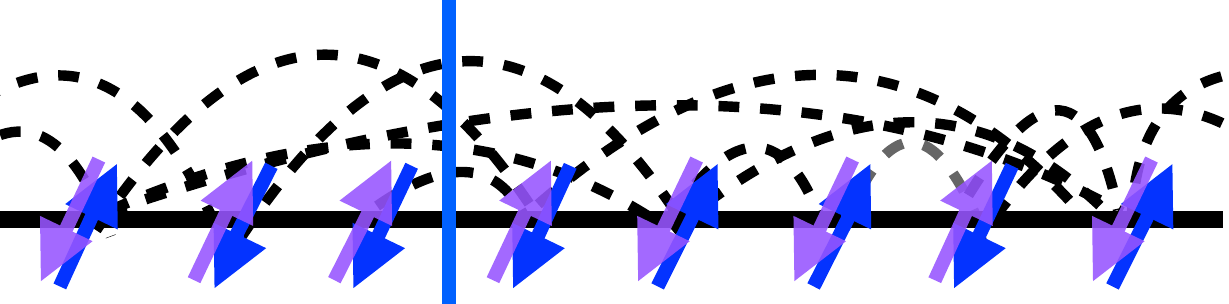}
 \end{center}
\caption{Schematic: entanglement across a cut through bond at position $x$.}
 \label{spinchaincutfig}
\end{figure}

Consider a chain with open boundary conditions that is in a pure quantum many-body state, and let $S(x,t)$ be the von Neumann entanglement entropy across a cut through the bond at position $x$ at time $t$. Formally,
\be
S(x,t) = - \Tr \rho_A(t) \log \rho_A(t)
\ee
where the subsystem $A$ contains the degrees of freedom to the left of $x$ (Fig.~\ref{spinchaincutfig}).  We could also consider a mixed state of the full system, in which case this is the von Neumann entropy of the subsystem to the left of the cut.

We will model the dynamics of $S(x,t)$ as  deterministic surface growth. In the next subsection we will motivate this using a toy model for entanglement growth in the presence of weak links, but first we describe the consequences for the coarse-grained dynamics.

The surface growth picture takes into account  two crucial physical constraints. First, the growth rate at a weak link $i$ is constrained by the local rate $\Gamma_i$. Second, the spatial slope of the entropy profile $S(x,t)$ is  constrained by the density of active degrees of freedom in the spin chain:
\ba
\partial S(x_i, t)/\partial t  &\leq \Gamma_i,
&
|\partial S(x,t)/\partial x|  &\leq s_\text{eq}.
\end{align}
Here $s_\text{eq}$ is the entropy density of the thermal state to which the system is locally equilibrating.  The second inequality follows from subadditivity of the von Neumman entropy, together with the assumption that \textit{local} reduced density matrices (for $O(1)$ adjacent spins) thermalize at late times.  In a lattice spin model at infinite temperature, $s_\text{eq}$ is the logarithm of the local Hilbert space dimension per unit length; this version of the inequality follows rigorously from subadditivity.  In the following we will rescale $x$ so that the inequality becomes
\be\label{slope_inequality}
|\partial S(x,t)/\partial x|  \leq 1.
\ee

If the system has any conserved densities, such as energy, that have a spatial distribution that is away from thermal equilibrium, then these equations for entanglement production are coupled to the transport equations for the conserved densities, and $\Gamma_i$ and $s_\text{eq}$ in general depend on the local densities.  But we will assume that any such conserved densities are close to equilibrium and it is only the entanglement that is out of equilibrium, as is natural in many global quenches.

As a result of these inequalities, each weak link $i$ imposes the constraint ${S(x,t)\leq S(x_i,0) + \Gamma_i t + |x-x_i|}$. We propose that at late times the entropy is essentially as large as it can be given these constraints:
\be\label{coarse-grained growth rule}
S(x,t) = \min_i \left\{S(x_i,0) + \Gamma_i t + |x-x_i| \right\}.
\ee
The entanglement across a bond at $x$ is therefore determined by a single locally `dominant' weak link. The spatial boundaries are taken into account by treating them as weak links with $\Gamma=0$.  For a pure state, the final profile at asymptotically late times, once the system has fully thermalized, is the pyramid $S(x,t) = \min\{ x, L-x \}$. These formulae of course neglect subleading corrections (see Sec.~\ref{sec_randomcircuit}); for example we know that the entanglement near the centre of the chain will depart from the maximal value by an $O(1)$ correction even as $t\rightarrow \infty$ \cite{Page1993}.

Fig.~\ref{growth_schematic} shows growth according to the rule in Eq.~\ref{coarse-grained growth rule}. In the next section we will see how it emerges at large length and time scales from a semi-microscopic toy model.

A simple optimization argument tells us how $S(x,t)$ scales with time if we start from a pure product state with $S(x,0)=0$.  Let $D$ be the typical distance to the dominant weak link at time $t$.  The weakest link within this distance scale will have a rate of order ${\Gamma_\text{min}\sim D^{-1/(a+1)}}$, so the two terms in $\Gamma_i t + |x-x_i|$ will scale as $D^{-1/(a+1)} t$ and $D$ respectively.  Minimizing with respect to $D$ gives
\ba\label{S_growth_eq}
S(x,t) & \sim t^{1/z_{S}}, &
z_S & = \f{a+2}{a+1}.
\end{align}
The dynamical exponent $z_S$ sets the typical lengthscale for entanglement at time $t$: for example the typical time for the entanglement profile to saturate in a finite system of size $L$ will be of order $L^{z_S}$.  It also governs the typical distance to the locally dominant weak link,
\be\label{typical_distance}
D \sim t^{1/z_S}.
\ee
This is the lengthscale for the dynamical coarsening of the entanglement pattern visible in Fig.~\ref{growth_schematic}.

A key feature of Eq.~\ref{S_growth_eq} is that the entropy grows `sub-ballistically' even for arbitrarily weak disorder: i.e. $z_S$ exceeds one even for arbitrarily large $a$. This is despite the fact that operator growth is ballistic for large enough $a$, as we will explain in Sec.~\ref{operator_spreading_sec}.  Eq.~\ref{S_growth_eq} differs from earlier results which effectively assumed the timescale for entanglement saturation of a large chain was equal to a timescale associated with the weakest link in the chain\cite{VoskHuseAltmanMBL,PotterVasseurParameswaranMBL}.  These previous works took a viewpoint of the `spreading' of entanglement, while we now argue that this process is more accurately viewed as the constrained local {\it production} of entanglement.

Since $S$ in Eq.~\ref{S_growth_eq} is the minimum of a set of uncorrelated random variables it is straightforward to find its full probability distribution in the limit of an infinite chain initiated in a pure product state.  If the density of weak links is $\rho$ and the distribution of local rates is (\ref{prob_gamma}) at small $\Gamma$, the cumulative probability distribution at long time is
\ba\label{entropy_distribution}
P\lf \text{entropy} > S \ri &=  \exp \lf
- c
 \f{S^{a+2}}{t^{a+1}}
\ri
\end{align}
with (restoring the equilibrium entropy density $s_\text{th}$)
\be
c=\f{2 A \rho }{s_\text{th} (a+1)(a+2)}.
\ee

The surface growth picture is restricted to the entanglement across a single cut; to generalize to regions with multiple endpoints, or to periodic boundary conditions, we may use the `directed polymer' description in Sec.~\ref{directed_polymer_sec}.

\subsection{Random circuit model}
\label{sec_randomcircuit}

Quantum circuit dynamics, with randomly chosen 2-site unitaries, capture many universal features of entanglement growth in translationally invariant systems \cite{EntanglementRandomUnitary}. This setup is easily adapted to give a toy model for entanglement growth in the presence of weak links.  This leads to a concrete surface growth problem defined on the lattice, from which the coarse-grained rule (\ref{coarse-grained growth rule}) can be seen to emerge in the scaling limit.  This is not a derivation of the surface growth picture for isolated 1D systems with time-independent Hamiltonian or Floquet dynamics, whose microscopic dynamics are very different from the toy model; however it motivates the surface growth picture, some of whose predictions can subsequently be checked by other means.

\begin{figure}[t]
 \begin{center}
  \includegraphics[width=\linewidth]{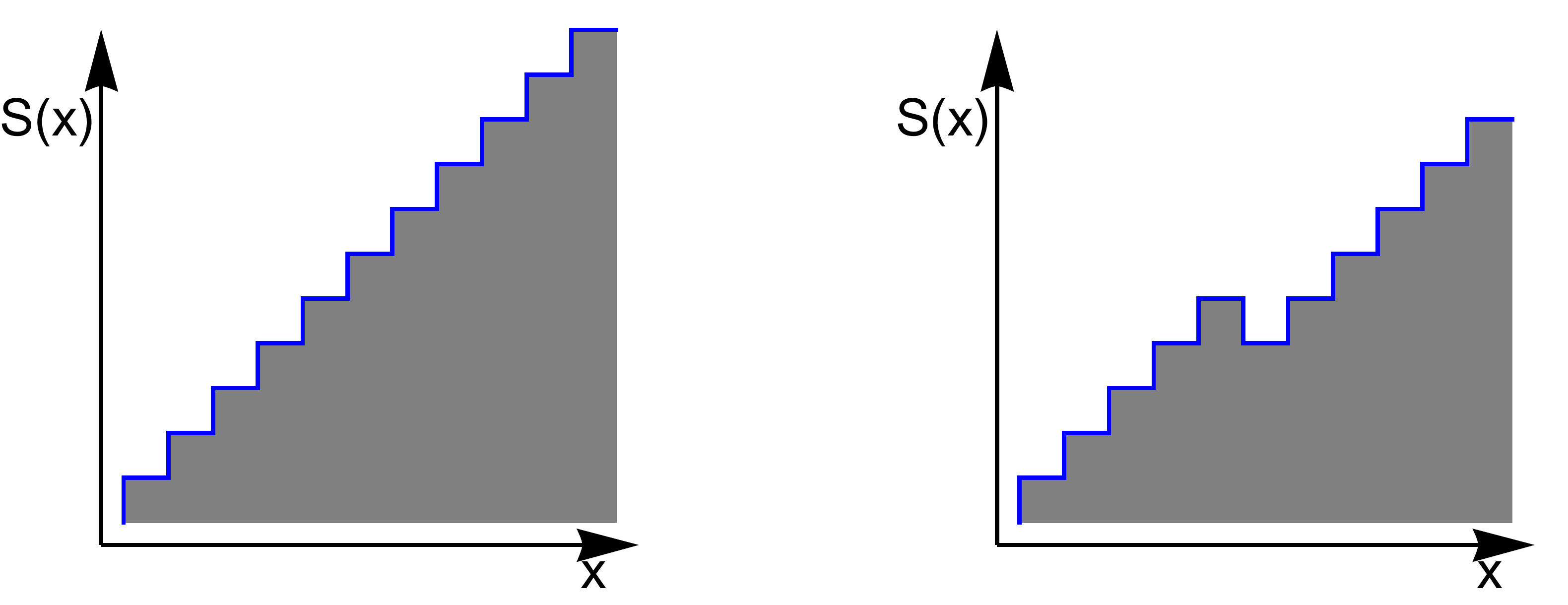}
 \end{center}
\caption{Left: maximum possible slope for the entanglement profile in the quantum circuit model, $\partial S/\partial x = 1$. Right: a staircase with a defect. After coarse-graining, a finite density $\rho$ of defects results in a slope $\partial S/\partial x = 1-2\rho$.}
 \label{stairconfig}
\end{figure}

\begin{figure}[b]
 \begin{center}
  \includegraphics[width=0.55\linewidth]{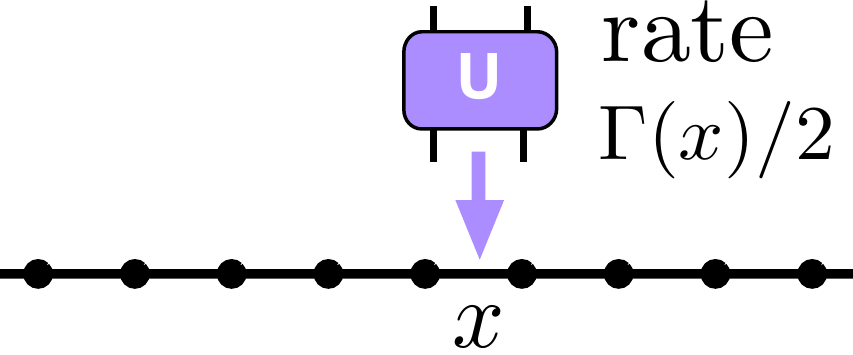}
 \end{center}
\caption{{Random circuit model: each bond $x$ receives  Haar-random unitaries at its own rate $\Gamma(x)/2$.}}
 \label{application_of_unitary}
\end{figure}

Consider a chain of `spins', each with a large local Hilbert space dimension, $q\gg 1$. The chain is initially in an unentangled product state. Random 2-site unitaries are then applied to adjacent spins in a Poissonian fashion, at rates $\f{1}{2} \Gamma(x)$ that depend on the bond $x$ (Fig.~\ref{application_of_unitary}). These rates are distributed at small $\Gamma$ as in Eq.~\ref{prob_gamma}.

At large $q$ the dynamics of the entanglement $S(x,t)$ across bond $x$ maps exactly to a classical surface growth model \cite{EntanglementRandomUnitary}. We absorb a factor of $\log q$ into the definition of $S$ (by defining the von Neumann entropy using a logarithm base $q$).  The entanglement $S(x,t)$ then
obeys a very simple dynamical growth rule:  Each time a unitary is applied to bond $x$, $S(x,t)$ increases to the maximal value allowed by the general constraint that  $S(x,t)$ can exceed $S(x\pm 1,t)$ for the neighbouring bonds by at most $1$. With this growth rule, the differences between adjacent heights are always $\pm 1$ at late times \cite{EntanglementRandomUnitary}.

The resulting dynamics is microscopically stochastic. However we may neglect the noise-induced fluctuations, since they are negligible in the long-wavelength limit.

First take $\Gamma$ to be constant (no weak links), and consider the growth of a region whose  coarse-grained slope $\partial S/\partial x$ is constant. If the surface is flat, $\partial S/\partial x = 0$, the growth rate is $\Gamma/4$ \cite{MeakinRamanlalSanderBall1986}. The important regime for us will instead be where the slope $|\partial S/\partial x|$ is  close to the maximal value of unity. Microscopically this means that the surface is close to the perfect staircase configuration shown in Fig.~\ref{stairconfig}, Left. Note that the growth rate vanishes in this configuration. However when $|\partial S/\partial x|$ is slightly smaller than unity, there is a small density $\rho$ of `defects' in the staircase, involving a local minimum of the height: see  Fig.~\ref{stairconfig}, Right.  These defects allow growth. Each time a unitary hits a defect, the local height increases by two units. The coarse-grained growth rate is thus $\partial S/ \partial t \simeq \rho \Gamma$. The coarse-grained slope is given by $|\partial S/\partial x|  = 1 - 2\rho$, so
\be\label{slope_dependence_eq}
\f{\partial S}{\partial t} \simeq \f{\Gamma}{2} \lf 1 - \left| \f{\partial S}{\partial x} \right| \ri.
\ee
Note that, microscopically, each time a unitary hits a defect the defect moves up the staircase by one step, so  defects run up the staircase at an average speed $\Gamma/2$.  The growth rate $\partial S/\partial t$ can be also be thought of as (twice) the `current' of these defects.

\begin{figure}[t]
 \begin{center}
  \includegraphics[width=0.95\linewidth]{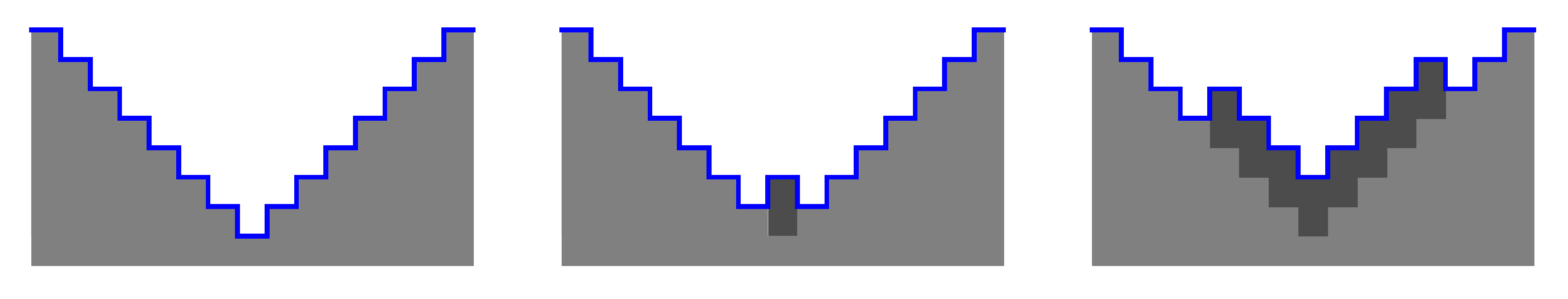}
 \end{center}
\caption{Growth at a weak link location (random circuit model). Unitaries are applied at the weak link  at a small rate $\Gamma_1$. Each such event increases the height at the origin by two units. Two  `defects' then travel up the sides of the double staircase at an average speed $\Gamma/2$.}
 \label{growth_at_weak_link}
\end{figure}

\begin{figure}[b]
 \begin{center}
  \includegraphics[width=0.49\linewidth]{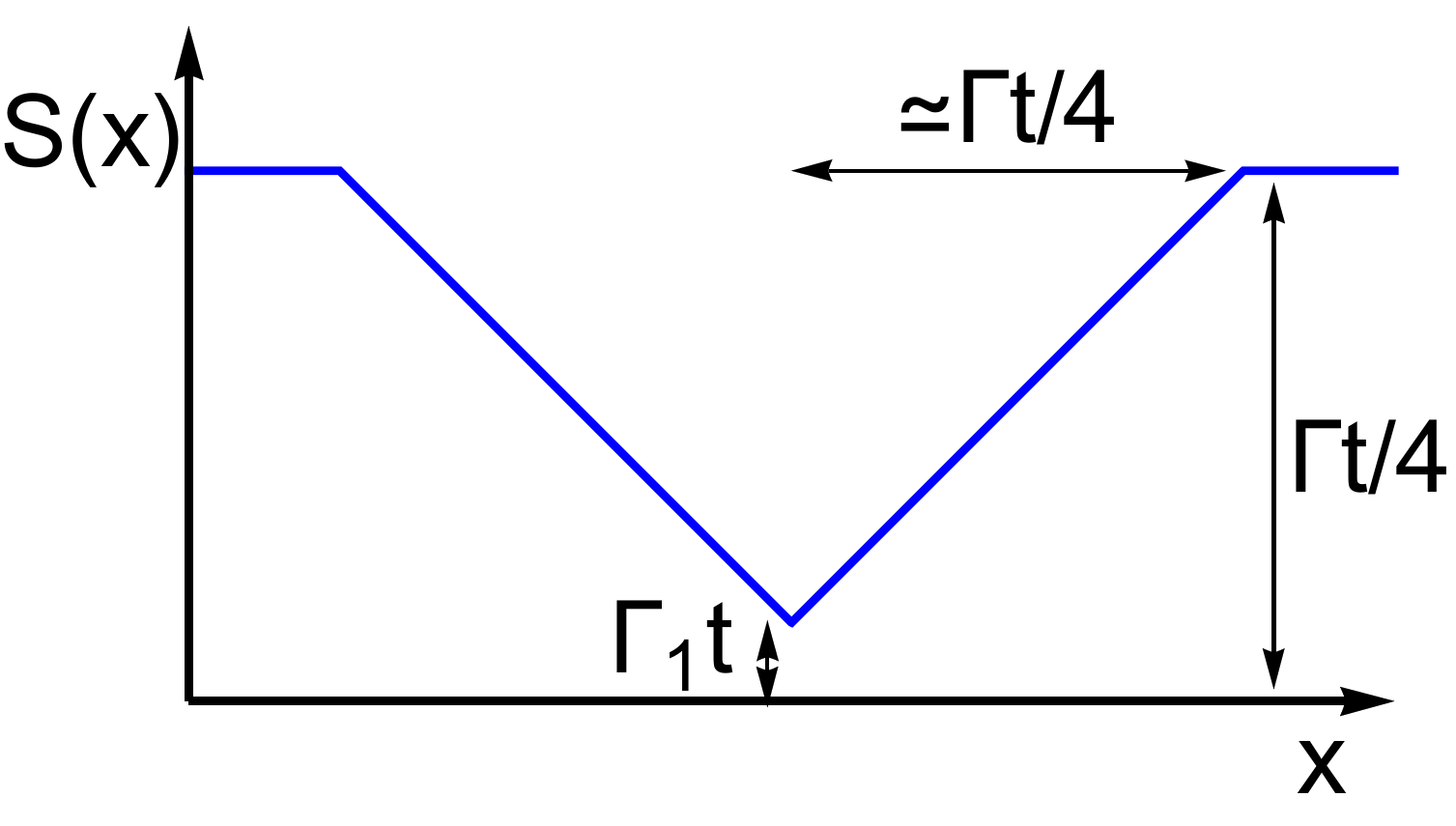}
  \includegraphics[width=0.49\linewidth]{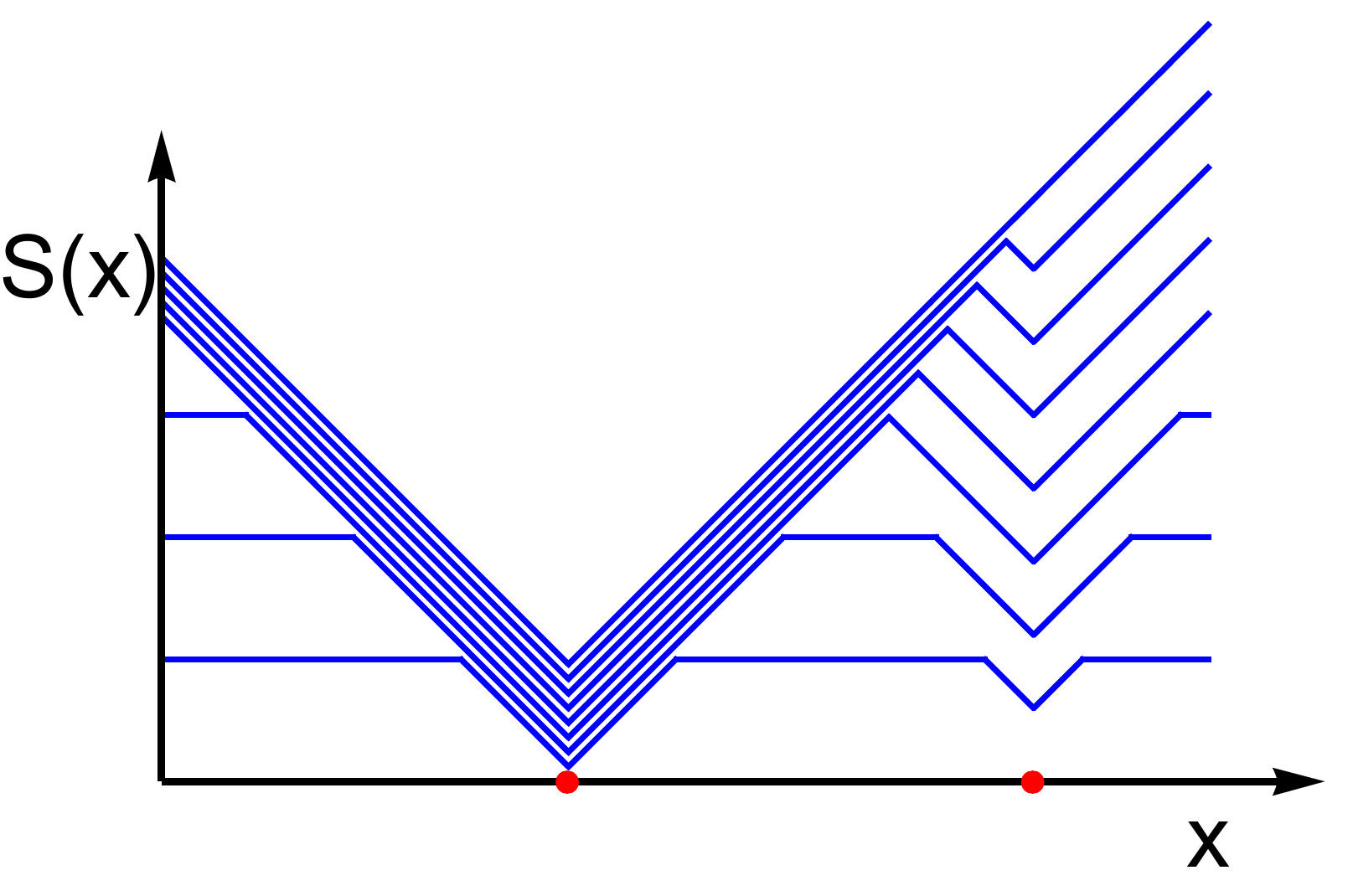}
 \end{center}
\caption{Left: Entanglement growth around a weak link characterized by a small rate $\Gamma_1$.  Right: Entanglement growth around a pair of weak links, showing how the weaker link dominates at late times. The figure shows eight equally spaced times.}
 \label{growth_at_weak_link_2}
\end{figure}

Next consider entanglement growth in a system with a single weak link at the origin, with  rate $\Gamma_1 \ll \Gamma$. At a typical time $t$ the local configuration resembles Fig.~\ref{growth_at_weak_link}, Left. Microscopically the weak bond is almost always a local minimum of the height profile. Unitaries are applied there at a rate $\Gamma_1/2$, causing growth at rate $\Gamma_1$. Each such event launches  one defect up each  staircase on the two  sides of the weak link. The growth rate of the adjacent regions is therefore set by the growth rate at the weak link, $\partial S/\partial t = \Gamma_1$.   The coarse-grained slope of the adjacent regions is fixed using (\ref{slope_dependence_eq}):
$\left| {\partial S}/{\partial x} \right| = 1 - {2\Gamma_1}/{\Gamma}$. For small $\Gamma_1$, the deviation of the slope from unity is small. Neglecting this deviation,
\be\label{1_weak_link_fmla}
S(x,t) \simeq \min \Big\{  \Gamma_1 t + |x| ,  \,  \Gamma t / 4 \Big\}~,
\ee
when the initial state is a pure product state.  This profile is shown in Fig.~\ref{growth_at_weak_link_2}, Left. The weak link influences a region of size  $\simeq \Gamma t/4$ on either side.

It is straightforward to generalize to multiple weak links. Consider two with rates $\Gamma_1$ and $\Gamma_2$ that are separated by a distance $l$. (Fig.~\ref{growth_at_weak_link_2}, Right.) We take $\Gamma_1 < \Gamma_2 \ll \Gamma$.  The regions of influence of the two weak links meet at a time $\simeq 2l/\Gamma$, giving a profile with a central peak.  The region of influence of the weaker link then gradually expands at the expense of the stronger link. At  time $t\simeq  l /(\Gamma_2 - \Gamma_1)$ the central peak hits the  link with the  rate $\Gamma_2$ and disappears. Subsequently the link with the larger rate has no effect on the coarse-grained configuration, being `dominated' by the weaker link.  (The link with rate $\Gamma_2$ does not affect the slope, or equivalently the density of defects, in this regime, except precisely at its location: the flow of defects is limited only by the slower rate $\Gamma_1$.) Again we may write
\be
S(x,t) \simeq \min \Big\{ \Gamma_1 t + |x-x_1|, \,   \Gamma_2 t + |x-x_2| ,  \,  \Gamma t/4   \Big\}.
\ee
The same logic extends to arbitrary numbers of weak links. The $\Gamma t/4$ term can be dropped since at large times every point $x$ is within the domain of influence of some weak link. This gives Eq.~\ref{coarse-grained growth rule}.

We may also quantify the subleading corrections to Eq.~\ref{coarse-grained growth rule}: the average gradient $|\partial S/\partial x|$ of the straight sections\footnote{
For large $t$ the profile consists of staircases of typical length $D\sim t^{1/z_S}$ with almost-maximal coarse-grained $|\partial S/\partial x|$. The minima between such staircases are weak links with typical strength $\Gamma_\text{min}\sim D^{-1/(a+1)}$  (see above Eq.~\ref{S_growth_eq}). The gradient at position $i$ is less than the maximum by $O(\Gamma_\text{min}/\Gamma_i)$. Summing this, the total height of the staircase is reduced from that of a perfect staircase by  $\Delta S$ of order $\Gamma_\text{min} \sum_{i=1}^D\Gamma_i^{-1}$ (note ${\Gamma_i\geq \Gamma_\text{min} + x/t}$) which gives the scaling in the text.
} is reduced from the maximal value of unity by an amount of order $D^{-1/(a+1)}$ when $a>0$ and of order $D^{-1}$ when $a<0$. Since $D\sim t^{1/z_S} \gg 1$ these corrections are indeed small.

\subsection{`Minimal cut' interpretation}
\label{directed_polymer_sec}

There is a general relationship between surface growth in 1+1D and the statistical mechanics of a directed polymer in a two-dimensional environment \cite{kpz}. The results discussed above may also be understood in this language, and this allows them to be generalized to more complex geometries.

In the context of entanglement the directed polymer may be viewed as a coarse-grained `minimal cut' through a unitary circuit representing the dynamical evolution \cite{EntanglementRandomUnitary}. We briefly summarize the main features of this coarse grained picture as it applies to the random circuit model. In the present case, with weak links, we obtain a directed polymer subject to pinning by vertical defect lines \cite{krug1993directed}.

\begin{figure}[t]
 \begin{center}
  \includegraphics[width=\linewidth]{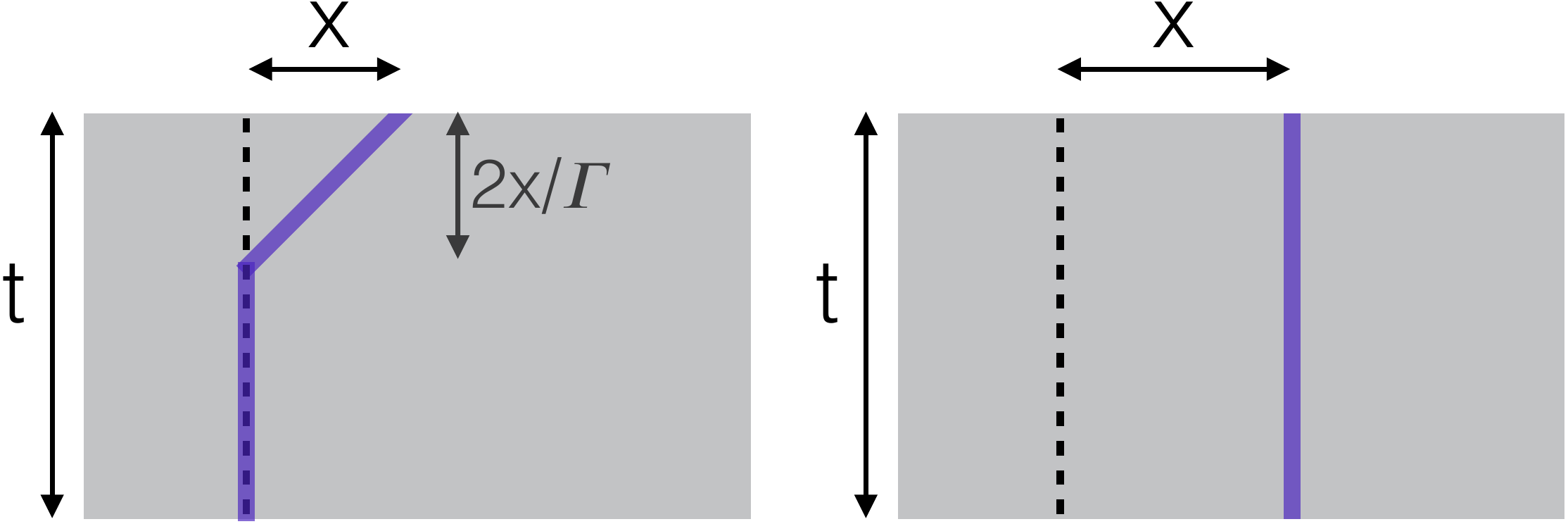}
   \end{center}
\caption{Minimal cut configurations determining $S(x,t)$ in an infinite system with a single weak link of strength $\Gamma_1$. The grey patch represents the unitary circuit and the thick line represents the (coarse-grained) minimal cut. Left: for ${x<\f{\Gamma t}{4} \f{(1-4\Gamma_1/\Gamma)}{(1-2\Gamma_1/\Gamma)}}$. Right: for ${x>\f{\Gamma t}{4} \f{(1-4\Gamma_1/\Gamma)}{(1-2\Gamma_1/\Gamma)}}$}
 \label{minimal_cut_schematic}
\end{figure}

The entanglement $S(x,t)$ is given by the `energy' of a minimal--energy cut which splits the space-time slice into two disconnected pieces --- see Fig.~\ref{minimal_cut_schematic}. (In this section we treat the time $t$ as a spatial dimension.) One endpoint of this cut must be at position $x$ on the top boundary, and the cut must disconnect the parts of the top boundary to the left and right of $x$. In an infinite system this means that the other endpoint of the cut must be at the bottom boundary. In the absence of weak links, the minimal energy such cut is vertical, and the energy per unit height is $\Gamma/4$ (in the notation of Ref.~\cite{EntanglementRandomUnitary} this is the entanglement rate $v_E$). In a finite system, the cut can terminate on the left- or right-hand spatial boundary.

To begin with it is sufficient to consider only horizontal and vertical cuts. The energy of a horizontal cut is equal to its extent in the $x$ direction. If we consider $S(x,t)$ in a clean semi-infinite system with a boundary at position 0, the minimal cut is vertical and of energy $\Gamma t /4$ for early times, while at late times the horizontal cut with energy $x$ is favourable; this gives {$S(x,t) = \min\{ \Gamma t/4, x\}$}.

A single weak link corresponds to a vertical defect line where the energy density per unit height is reduced and equal to $\Gamma_i$. For large $t$, it is worthwhile for the polymer to travel a large horizontal distance to take advantage of this favourable energy density. It must of course `pay' in energy for the non-vertical section required to reach the defect. For a crude picture we can consider only horizontal and vertical segments; it is easy to see that in an infinite system we then recover Eq.~\ref{1_weak_link_fmla}.

In more detail, the energy of a segment of horizontal extent $x$ remains equal to $x$ even if the segment is at a finite angle to the horizontal, so long as this angle is small enough\footnote{This can be seen from the microscopic picture of the polymer as a `minimal cut' through the large-$q$ unitary circuit. The energy of the polymer is equal to the number of bonds it cuts. A horizontal cut of length $x$ cuts $x$ bonds. This horizontal cut can be deformed to one with a finite coarse-grained slope, with the same energy, so long as the slope is $\geq 2/\Gamma$ (this is the typical vertical distance that the cut can travel before being blocked by a unitary).} (this is true for slopes $\geq 2/ \Gamma$). This means that the true minimal cut configuration for small $x$ is as shown in Fig.~\ref{minimal_cut_schematic}. This corresponds to taking subleading corrections in the slope of $S(x)$ into account, and gives ${S(x,t) = \Gamma_1 t + (1-2\Gamma_1/\Gamma) |x|}$ in agreement with the previous section.

This picture may be extended to multiple weak links. The typical transverse excursion for a given cut is of order $D\sim t^{1/z_S}$, so much smaller than $t$ (Eq.~\ref{typical_distance}). It may also be extended to a region with multiple endpoints. Consider the entanglement of a finite region of length $x$ in an infinite system. At early times the minimal cut configuration involves two disconnected cuts, giving an entanglement of order $t^{1/z_S}$ which is the sum of two independent random variables distributed as in Eq.~\ref{entropy_distribution}. Once this becomes equal to $x$, a configuration with a single horizontal cut becomes favourable.

\section{Operator spreading}
\label{operator_spreading_sec}

Heisenberg time evolution will  transform a local operator, for example the Pauli matrix $X_0$ located at the origin in a spin chain, into a  complex object $X_0(t) = U(t)^\dag X_0 U(t)$ which acts nontrivially on many sites. The spatial extent of this growing operator may be quantified using the commutator with a local operator at site $x$ \cite{LiebRobinson}. In particular one can define the recently much-studied object \cite{larkin, kitaev, shenker_stanford_1, shenker_stanford_2,Roberts_Shocks,maldacena2016,Aleiner, Sachdev_CriticalFermiSurface,  chowdhury2017onset, Sachdev_DiffusiveMetals, 
 Stanford, Altman, gu2016local, roberts2016lieb, HuangMBLOTOC,FanMBLOTOC,YuChenMBLOTOC,SwingleMBLOTOC,HeMBLOTOC,ChenMBLOTOC,SlagleMBLOTOC,roberts2015diagnosing,dora2016out,bohrdt2016scrambling,luitz2017information,leviatan2017quantum}
\be\label{OTO_definition}
C(x,t) = - \f{1}{2}  \Tr \rho_\beta [X_0(t), X_x]^2,
\ee
where $\rho_\beta$ is the density matrix $e^{-\beta H}/Z$ at the appropriate temperature. Expanding the squared commutator gives the `out-of-time-order' (OTO) correlator
\be
C(x, t) = 1 -  \Tr \rho_\beta X_0(t) X_x X_0(t) X_x.
\ee
$C(x,t)$ is of order one in a spatial region whose size grows with $t$, and $C(x,t)$ vanishes far outside this region.  In translationally invariant systems the size of the operator, as measured by $C$, grows ballistically with a speed $v_B$ known as the butterfly speed.

\begin{figure}[t]
 \begin{center}
  \includegraphics[width=0.8\linewidth]{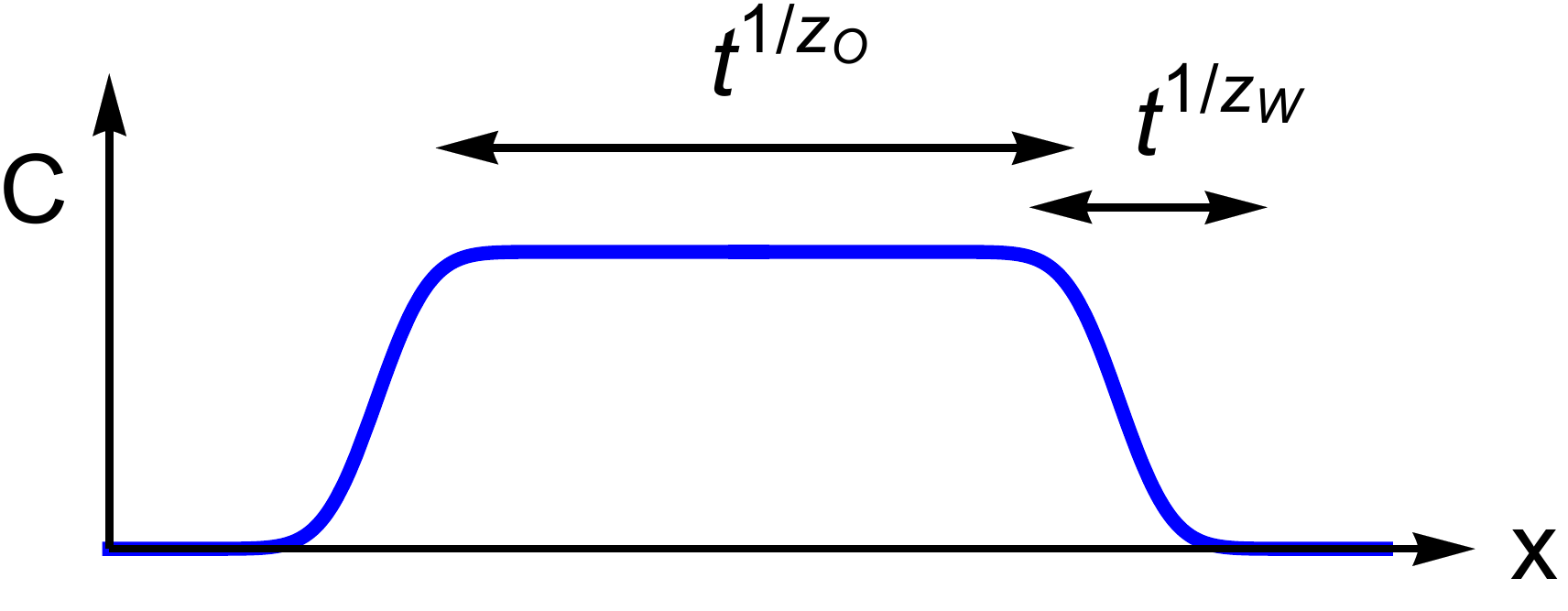}
   \end{center}
\caption{Definition of dynamical exponents $z_O$ and $z_W$ governing the size of a spreading operator and the size of the front [as measured using the OTO correlator $C(x,t)$, Sec.~\ref{operator_spreading_sec}]. The front is only well-defined in the weak disorder regime ($a>1$) where $z_O=1$ and $z_W < 1$.}
 \label{OTO_schematic}
\end{figure}

Here we address the size and `shape' of spreading operators in the thermal Griffiths phase; see Fig.~\ref{OTO_schematic}. For simplicity we consider the case $\beta = \infty$, but we do not expect this to change the basic results. We determine dynamical exponents $z_O$ and $z_W$ which give respectively the typical size $t^{1/z_O}$ of the operator at time $t$ and the typical size  $t^{1/z_W}$ of its `front'  --- the region in which $C$ is of order one, but smaller than the saturation value. The shape of the operator turns out to be qualitatively different for weak and strong disorder. For weak disorder $z_W > z_O$: in this regime the front is much smaller than the `plateau' region in which  $C$ has saturated to its maximum value. Conversely for strong disorder $z_W=z_O$, so that $C$ does not have a well-defined plateau when lengths are scaled by the spreading length.

Our starting point is a picture for operator spreading in 1D systems developed on the basis of calculations in random circuits in Ref.~\cite{OperatorGrowth}  (see also parallel work which appeared recently \cite{CurtPaper}). It is shown there that for $x>0$ and for long times we may write (we switch to a continuum notation)
\be\label{C_rho_relation}
C(x,t) =  \int_x^\infty \dd x' \rho(x', t),
\ee
where $\rho(x,t)$ is a conserved density
\be
\int_0^\infty \rho(x,t) = 1,
\ee
and where ---  in a random circuit without weak links --- $\rho(x,t)$  behaves essentially like the probability density of a random walker with a bias in favour of rightward steps. The average speed $v_B$ of the walker sets the butterfly velocity.  For $x < v_B t$, the bulk of the density is within the range of integration in (\ref{C_rho_relation}), yielding $C(x,t)\simeq 1$; for $x>v_B t$ the density is mostly outside the range of integration, giving $C(x,t)\simeq 0$. The transition region broadens diffusively, with width $D\sqrt{t}$. This picture in terms of the density $\rho$  generalizes very naturally to the situation with weak links, where $v_B$ is no longer necessarily nonzero.

Let us briefly summarize the meaning of the density $\rho$ \cite{OperatorGrowth}. First write the spreading operator at time $t$ in the basis of products of Pauli matrices,
\be\label{string_expression}
X_0(t) = \sum_{\mathcal{S}} a_\mathcal{S}(t) \, \mathcal{S}.
\ee
Here $\mathcal{S}$ is a string (product) of Pauli matrices at different sites. These strings satisfy
\be
\Tr \rho_\infty \mathcal{S} \mathcal{S}' = \delta_{\mathcal{S}\mathcal{S}'},
\ee
and since  $\Tr \rho_\infty X_0(t)^2 = 1$ we have ${\sum_\mathcal{S} a_\mathcal{S}(t)^2 = 1}$.  The density $\rho(x,t)$ is the `fraction' of strings which \textit{end} at position $x$ (a string $\mathcal{S}$ ends at $x$ if $x$ is the rightmost site at which it acts nontrivially):
\be
\rho(x,t) = \sum_{\substack{\mathcal{S}\\\text{(ends at $x$)}}} a_\mathcal{S}^2.
\ee
The density $\rho(x,t)$ is evidently conserved and is normalized to one, $\sum_x \rho(x,t) = 1$, despite the fact that the \textit{number} of distinct strings contributing to this density grows exponentially with time.

Consider $C(x,t)$ for $x>0$. Inserting the expression (\ref{string_expression}) shows that $C(x,t)$ is
\be
C(x,t) = 2  \, {\sum_{\mathcal{S}} }'a_\mathcal{S}^2,
\ee
where the primed sum includes only those strings whose commutator with $X_x$ is nontrivial. Strings whose right endpoint is at a position to the left of $x$ cannot contribute to this sum, but an $O(1)$ fraction of those whose right endpoint is to the right of $x$ do contribute. Specifically we expect\footnote{A given string can either have $X, Y, Z$ or $1$ at site $x$ (two of which commute with $X$), and we expect all options to be equally likely \cite{OperatorGrowth} deep in the interior of the operator.} this fraction to be $1/2$, giving (\ref{C_rho_relation}).  In a random circuit without weak links, $\rho(x,t)$ may be argued to satisfy a noisy diffusion equation with bias \cite{OperatorGrowth}.

This motivates the following picture for a system whose dynamics is deterministic  but spatially random.  We first consider the spreading of an operator through a single weak link, and then generalize to a Griffiths chain with many weak links.

\subsection{Operator spreading across one weak link}

\begin{figure}[t]
 \begin{center}
  \includegraphics[width=\linewidth]{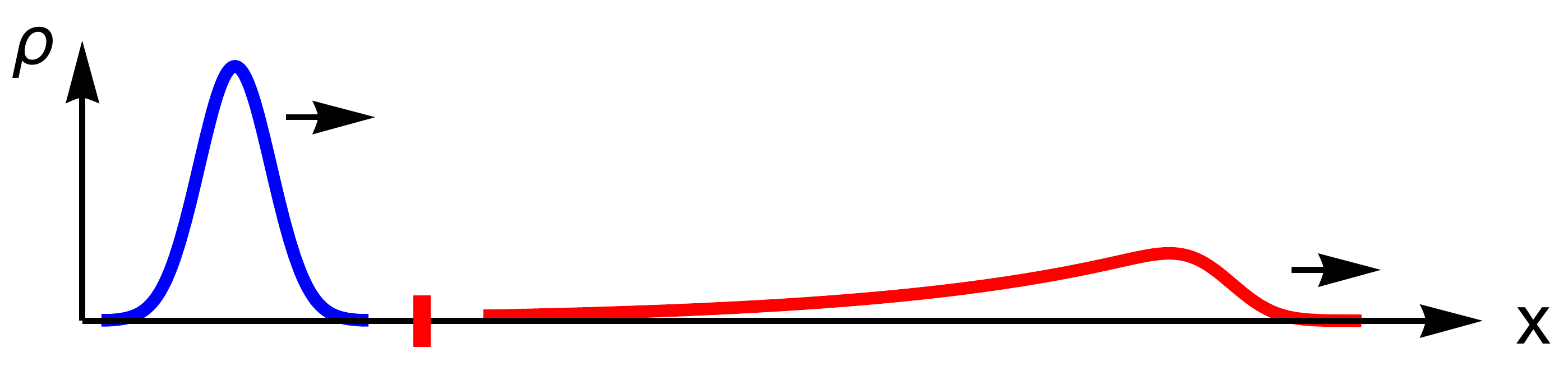}
    \includegraphics[width=\linewidth]{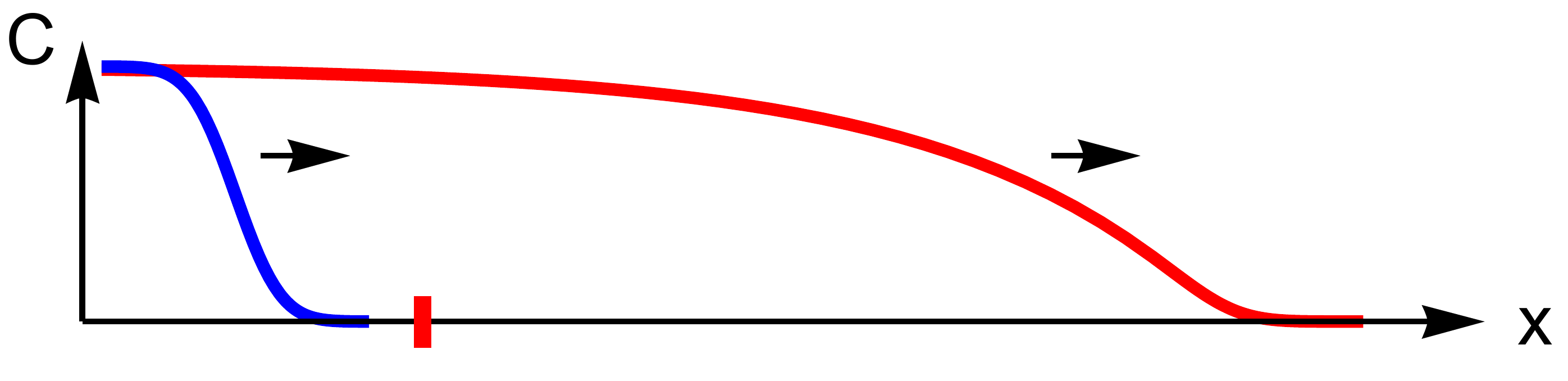}
 \end{center}
\caption{Operator spreading caused by passage through a weak link (schematic). Top: the spreading of the `wavepacket' $\rho(x,t)$ (Eq.~\ref{C_rho_relation}). Bottom: the corresponding spreading of the front of the commutator $C(x,t)$ (Eq.~\ref{OTO_definition}).}
 \label{weak_link_spreading}
\end{figure}

Consider two regions in which operators spread at speed $v_B$, separated by a weak link  characterised by a very small rate $\Gamma$.  For simplicity we think of the weak link as a weak bond at position $x=0$, and  consider its effect on the rightward front of an operator which is initially localized in the leftward part of the system.

First, the density $\rho(x,t)$ advances to the weak link, which we take to be located at position $x=0$, reaching the weak link at time $t_*$. We neglect the diffusive spreading in $\rho$ prior to passing through the weak link. It is easy to restore this effect in what follows by a more detailed treatment of biased diffusion on a chain with weak links, but it is not important for the scaling exponents below. The density then leaks through at a rate $\Gamma$, so
\be
\rho(0^-, t) = e^{-\Gamma(t-t_*)}
\ee
where $0^-$ is the lattice site to the left of the weak link. The density on the other side of the weak link is
\be
\rho(0^+,t) = \f{\Gamma}{v_B} e^{-\Gamma(t-t_*)}.
\ee
(Walkers hop across at rate $\Gamma$, and are whisked away at speed $v_B$.) Since we neglect spreading of $\rho$ within the homogeneous region, the density to the right of the weak link is related to the above by ${\rho(x,t)  = \rho(x-v_Bt)}$: for ${0< x < v_B(t-t_*)}$ we have
\ba
\rho(x,t) = \f{\Gamma}{v_B} \exp \lf  \f{\Gamma}{v_B} \left[  x - v_B (t-t_*) \right]  \ri.
\end{align}
At times  $\gtrsim 1/\Gamma$ (but small compared to $v_B^2/\Gamma^2 D$, when diffusive spreading becomes comparable) the packet $\rho(x)$ is of width $\sim v_B/\Gamma$ and of height $\Gamma/v_B$.  See Fig.~\ref{weak_link_spreading}, Top.

\subsection{Operator spreading across many weak links}

To understand spreading across multiple weak links it is useful to think of each one as performing a linear transformation on the `wavepacket' $\rho(x,t)$. We go into a frame moving at speed $v_B$. The foregoing tells us that if the initial wavepacket is
\be
\rho_0(x) = \delta(x-x_0),
\ee
then the new wavepacket is (we define $\gamma = \Gamma/v_B$):
\be
\rho_1(x) = \gamma e^{ \gamma ( x -x_0 )}  \qquad (x<x_0).
\ee
In other words, by linearity,
\be
\rho_1(x) = \gamma \int_0^\infty \dd y e^{-\gamma y} \rho_0(x+y).
\ee
This transformation may be iterated.\footnote{E.g. for one choice of initial condition, iterating with the same value of $\gamma$ gives $\rho_k = \gamma \f{(-\gamma x)^k}{k!} e^{\gamma x}$ ($x<0$), which becomes Gaussian for large $k$.} It preserves the normalization of $\rho$ and it acts in a simple way on the mean and variance. If the $k$th weak link encountered has strength $\gamma_k$,
\ba\notag
\< x\>_{k+1} &= \< x\>_k - \gamma_k^{-1},
\\
\<\< x^2 \>\>_{k+1}& = \<\< x^2 \>\>_{k}+ \gamma_k^{-2}.
\label{mean_and_variance_wavepacket}
\end{align}

\subsection{Operator spreading in the Griffiths phase}
\label{operator_spreading_griffiths_phase}

Next let us consider the leading edge of the commutator as it passes through a sequence of weak links $\Gamma_k$, with a probability distribution $P(\Gamma)\sim \Gamma^a$. (We take the separation of the weak links to be unity.) There are three separate questions: (i) How far has the leading edge travelled at time $t$? (ii) What is the typical {width} of the leading edge after a time $t$, \textit{within} a given sample, i.e. for a given realization of the quenched disorder? (iii) How much variation in the position of the leading edge is there {between} \textit{different} disorder realizations?

After traveling a distance $x$ the wavepacket has passed through $O(x)$ weak links. From the formula for the mean, the position of the wavepacket is
\be\label{front_position}
x \simeq v_B t - \sum_{k=1}^x \gamma_k^{-1},
\ee
where we have made the approximation that all parts of the wavepacket have passed through the same number of weak links. (This simplification does not change the scaling of $x$, but must be considered more carefully for the fluctuations below.) The scaling of the  sum in (\ref{front_position}) depends on the value of $a$. In the regime $a>0$, the sum is proportional to the number of terms, whereas for $a<0$ it is dominated by the smallest $\gamma$, which is of order $x^{-1/(a+1)}$. This gives:
\ba
a>0: & & x& \sim t, & z_{O} &= 1, \\
a<0:&  & x&\sim t^{a+1}, & z_{O} &= {1}/({a+1}),
\end{align}
where we have introduced the dynamical exponent $z_O$ governing the size of a spreading operator.

Now let's estimate the spreading of the wavepacket $\rho(x,t)$ within a given sample. The variance formula (\ref{mean_and_variance_wavepacket}) gives
\be\label{width_formula}
\text{width}^2 \sim  \sum_{k=1}^x \gamma_k^{-2}.
\ee
This formula will give the correct scaling in the regime where all parts of the wavepacket have passed through $O(x)$ links. This is the case for $a>0$, where (\ref{width_formula}) gives a width much smaller than $x$.  Interestingly, there are two distinct behaviours within the $a>0$ regime: for $a>1$ the sum is of order $x$, while for ${0<a<1}$ it is dominated by the minimal element:
\ba
a>1&: &  x&\sim t, &  \text{width} & \sim {t}^{1/2}, \\
0<a<1&: & x&\sim t, & \text{width}& \sim t^{1/(a+1)}.
\end{align}
When $a<0$, naive application of the variance formula gives $\text{width}\gg x$,  showing that the approximation that all parts of the wavepacket have travelled through $O(x)$ weak links breaks down.  In this regime we expect simply that $\text{width}\sim x$, i.e.
\ba
a<0&: & x& \sim t^{a+1}, & \text{width}& \sim t^{a+1}.
\end{align}
To see this, note that at time $t$ some of the wavepacket will have passed through the weakest nearby link, which is at distance $x\sim t^{a+1}$ and has rate $\gamma\sim 1/t$, but some of the wavepacket will still be held up by the second weakest nearby link, which is an $O(1)$ fraction of the distance away.

Putting these results together,
\ba
a>1&: &  x&\sim t, &  \text{width} & \sim {t}^{1/2}, \\
0<a<1&: & x&\sim t, & \text{width}& \sim t^{1/(a+1)}, \\
a<0&: & x& \sim t^{a+1}, & \text{width}& \sim t^{a+1}.
\end{align}
These formulas define a dynamical exponent $z_W$ governing the width of the front of a spreading operator.

So far we have considered the width within a given sample. Sample-to-sample variations in the front position are even more simply understood using (\ref{front_position}). We find that in all regimes they scale with the same power of $t$ as the width of the front within a given realization.

The exponents above are written in terms of the parameter $a$ governing the distribution of weak link timescales. In using the same value of $a$ for operator spreading and for entanglement spreading, we are making the natural (but unproven) assumption that the timescales for entanglement growth and operator spreading are of the same order for a severe weak link.

There is a relationship between the growth of the second Renyi entropy and the spreading of the operators appearing in an expansion of the reduced density matrix, which has been used to give heuristic pictures for entanglement growth \cite{ho2017entanglement, mezei2016entanglement}. Nevertheless the growth `speed' associated with entanglement is in general smaller than that for the spreading of operators, even in clean systems \cite{EntanglementRandomUnitary, mezei2016entanglement}. The Griffiths phase is an extreme example, where the two lengthscales grow with different powers of time.

To summarize, operators have a well-defined front and a nonzero butterfly velocity only when $a>0$. In all regimes the width of the front increases with time, and there is a change in the exponent governing this width at $a=1$.  The various dynamical exponents are summarized below in Table~\ref{dynamical_exponent_table}.

\subsection{Operator entanglement}

We may also consider the entanglement of a spreading operator, viewing the operator as a state in a ``doubled'' system \cite{jh}.  The growth of this operator entanglement within the region in between the two ``fronts'' of the spreading operator is governed by essentially the same physics as that governing the growth of the entanglement of states.  Thus we expect the operator entanglement to grow with dynamic exponent $z_S$.  Since $z_S>z_O$ for all allowed finite $a$ within $-1<a<\infty$, the entanglement $S_\text{op}$ across the midpoint of a spreading operator is less than ``volume-law'' as long as it is spreading, where we call the distance $\ell$ between the two fronts the operator's ``volume'': $S_\text{op} \sim \ell^{z_O/z_S}$. Interestingly, this exponent is non-monotonic. It is minimal at $a=0$ (assuming that timescales are characterized by a single exponent $a$), and the `volume law' exponent of unity is recovered in both limits, $a\rightarrow \infty$ and $a\rightarrow -1$. 

Once the operator reaches the ends of the chain it can then become volume-law entangled.

\section{Conserved quantities}
\label{conserved_quantities_sec}

Here we revisit the dynamics of conserved quantities in the Griffiths phase, considered previously in Refs.~\cite{VoskHuseAltmanMBL,PotterVasseurParameswaranMBL,AgarwalAnomalousDiffusion}, in order to compare with the spreading of operators and entanglement.  We recover the dynamical exponent found previously for conserved quantities:
\be\label{conserved_quantities_exponent}
z_C = \min \left\{ 2, \f{a+2}{a+1} \right\}.
\ee
In writing (\ref{conserved_quantities_exponent}) in terms of the exponent $a$ (Eq.~\ref{prob_gamma}) we have assumed that a weak link has a single associated timescale which governs both entanglement growth and `hopping' of conserved quantities across the weak link. If there are cases where this assumption fails,  a distinct exponent $a_C$ could appear in the formula (or even multiple $a_C$s for different conserved quantities).  It is  natural to expect at least that $a\leq a_C$: the `hopping' of (say) a conserved charge across the weak link will generically induce $O(1)$ entanglement, so the rate for entanglement production should not be parametrically smaller than that for conserved quantities.

Since the system locally looks thermal at late times, we may treat the dynamics of conserved quantities  as a classical random walk --- say on a 1D lattice, in continuous time --- which is `bottlenecked' by weak links. We treat the weak links as bonds where the hopping rate is small. This setup preserves detailed balance.

For weak disorder, the dynamics is diffusive. To see when diffusive scaling breaks down, consider two adjacent weak links with $\Gamma \leq \Gamma_0$. Their interior defines a box of typical size
\be\label{box_size}
\Delta \sim \Gamma_0^{-(a+1)}.
\ee
Now compare the time for a diffusing particle to cross this box, ${t_\text{diff} \sim \Delta^2}$, with the time for which the particle is detained by a single weak link of strength $\Gamma_0$. Once the walker reaches the weak link it must revisit it $O(1/\Gamma_0)$ times before it succeeds in hopping across. By standard diffusive scaling, the time required for this number of revisits is ${t_\text{traverse}\sim 1/\Gamma_0^2 \sim \Delta^{2/(a+1)}}$. When $a>0$ we have $t_\text{traverse} \ll t_\text{diff}$, and we expect diffusive scaling to be stable. On the other hand when $a<0$ the two weak links trap the walker inside the box for much longer than $t_\text{diff}$. In this regime $t_\text{traverse}$ will be of the same order as the time required to explore the box `ergodically'. The rate to cross one of the weak links is then the product of the fraction of the time spent adjacent to the weak link, namely $1/\Delta$, with the $O(\Gamma_0)$ rate at the weak link. The typical time required to escape the box is therefore
\be
t \sim \Delta/ \Gamma_0 \sim \Delta^{(a+2)/(a+1)}.
\ee
This gives the dynamical exponent quoted above. This exponent agrees with the random walk model of Ref.~\cite{PotterVasseurParameswaranMBL}, but the behaviour of trajectories is different.\footnote{In the regime $a<0$ a walker takes a time of order $L/\Gamma_\text{min}$ to traverse a sample of size $L$, where $\Gamma_\text{min}$ is the typical size of the weakest link in the sample. In the above model this timescale is associated with trajectories that traverse the weak link $O(1)$ times, whereas in the model of Ref.~\cite{PotterVasseurParameswaranMBL} it is associated with trajectories which traverse the weak link $O(L)$ times.}

The transition between diffusive and subdiffusive behaviour has been observed numerically in a disordered Heisenberg chain \cite{DiffusionHeisenbergChain}; see also Refs.~\cite{LuitzLaflorencieAlet, AgarwalAnomalousDiffusion, BarLevNoDiffusion, NoDiffusion1D}.

 If a quantum quench starts from a sufficiently inhomogeneous initial state, the relaxation of conserved quantities will affect the growth of entanglement. An extreme example is a 1D spin chain, with conserved $S_z$, which starts in a domain wall state  with fully polarized up spins on the left and fully polarized down spins on the right. Since a polarized region has trivial dynamics, entanglement can only be generated in the growing central region where the polarization has been destroyed. For an initial state with only \textit{short-range} correlated randomness in conserved quantities, the local expectation value of the conserved quantity will relax to equilibrium with fluctuations of order $t^{-1/2 z_C}$. These fluctuations will lead to fluctuations in the local entangling rates $\Gamma$ for the weak links, but these will be negligible at late times.

\section{Dynamical exponent summary}
\label{sec:exponent_summary}

The dynamical exponents we have found are summarized in Table.~\ref{dynamical_exponent_table}, under the assumption  (see caveats in Secs.~\ref{operator_spreading_griffiths_phase}, \ref{conserved_quantities_sec})  that the long timescales characterizing a weak link are distributed with the same power law.

\begin{table}[h!]
\begin{center}
\begin{tabular}{|c|c|c|c|c||}
\hline
 & $-1 < a < 0$  & $ 0 < a < 1$ & $1 < a$  \\
 \hline
$z_S$ & \multicolumn{3}{c|}{$(a+2)/(a+1)$}  \\
\hline
$z_C$ &  $(a+2)/(a+1)$ & \multicolumn{2}{c|}{$2$} \\
\hline
$z_O$ &  $1/(a+1)$ & \multicolumn{2}{c|}{$1$}  \\
\hline
$z_W$ & $1/(a+1)$ & $a+1$ & $2$  \\
\hline
\end{tabular}
\end{center}
\caption{Dynamical exponents governing lengthscales for entanglement growth ($z_S$); spreading of conserved quantities ($z_C$); spreading of quantum operators under Heisenberg time evolution ($z_O$); width of the `front' of a spreading operator ($z_W$). Here we assume a single exponent $a$ governs the distribution of timescales (Eq.~\ref{prob_gamma}) for the various processes at a weak link, see text.}
\label{dynamical_exponent_table}
\end{table}%

\section{Bounding entanglement growth across a Griffiths region}
\label{explicit_griffiths_argument}

In previous sections we assumed that a Griffiths region could be characterised by a local entanglement growth rate, $\Gamma$, which vanishes as the region becomes large. Here we show analytically that the rate for entanglement growth across a Griffiths region is exponentially slow when the length $\ell$ of the Griffiths region is large.

We will start by considering a trivial kind of weak link --- a weak bond in a spin chain. For this simple case, a standard rigorous result provides a bound on the entanglement growth rate. A Griffiths region is more complicated: although it acts as a weak link in a coarse-grained sense, it is not equivalent to a simple weak bond, and the degrees of freedom within the Griffiths region are strongly coupled. Nevertheless we show below that the bound can be extended --- nonrigorously --- to this case by making use of the `l-bit' picture for many-body localized systems \cite{PhenomenologyOfFullyMBL, SerbynLocalConservationLaws}.

\begin{figure}[t]
 \begin{center}
  \includegraphics[width=0.9\linewidth]{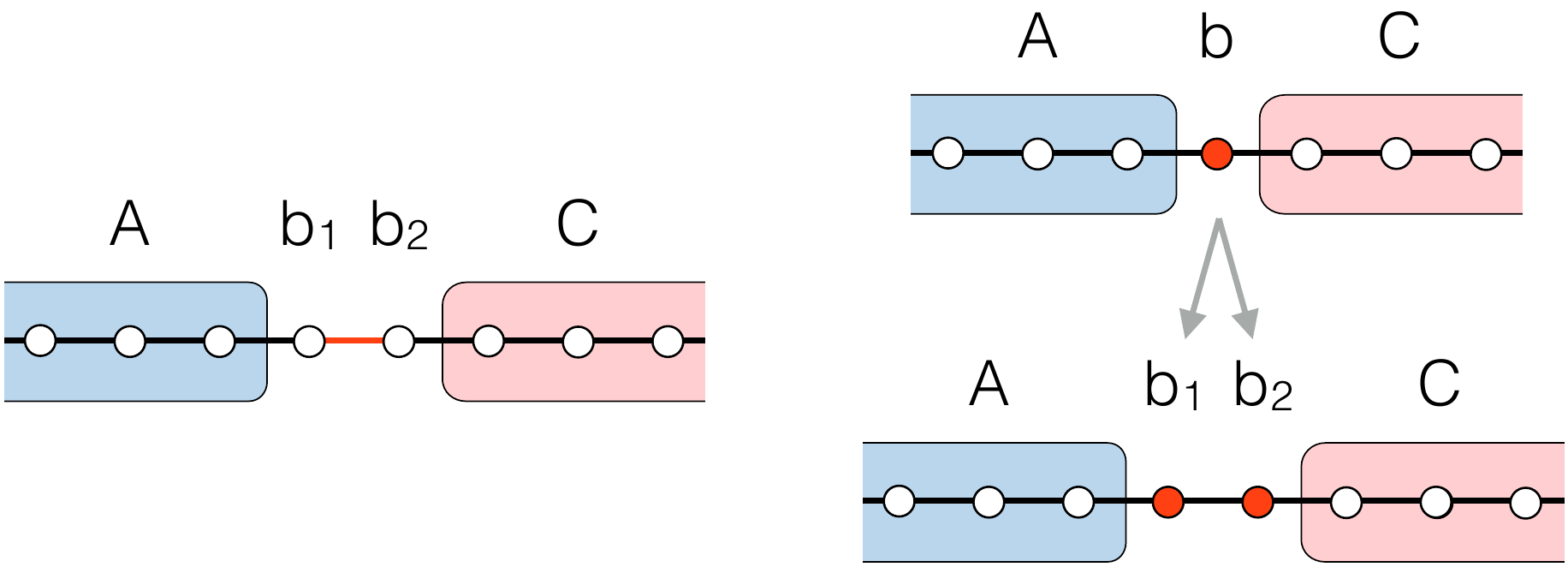}
 \end{center}
\caption{Two types of microscopic weak link.  Left: an infinite spin chain with a weak bond connecting spins $b_1$ and $b_2$. Right (upper): a  chain in which the $Z$ component of spin $b$ is almost conserved. This can be mapped to a chain with a weak bond by duplicating $b$ (lower).}
 \label{weak_link_types}
\end{figure}

First consider a spin chain in which one bond has a very small coupling. Label the degrees of freedom as in Fig.~\ref{weak_link_types}, Left. $b_1$ and $b_2$ denote the spins to the left and right of the weak bond respectively, and $A$ and $C$ contain the other spins on the left and the right respectively. The Hamiltonian may be written
\be
H = H_{Ab_1} + H_{b_1 b_2} + H_{b_2 C},
\ee
reflecting the fact that the two sides are coupled only via spins $b_1$ and $b_2$. For example, for an infinite chain with Ising interactions and longitudinal and transverse fields, we would have
\ba
H_{b_1 b_2} & = J_\text{weak} Z_0 Z_1, \\
H_{A b_1} & =  \sum_{i<0} \lf J Z_{i-1}Z_i +  h X_i + g Z_i \ri,
\\
H_{b_2 B} & = \sum_{i>1} \lf J Z_{i-1}Z_{i}+h X_i + g Z_i \ri,
\end{align}
with $J_\text{weak} \ll J$. This is a weak link of a simple kind: the systems $Ab_1$ and $b_2 B$ are coupled only by a small term in the Hamiltonian. In this simple situation there is a rigorous bound on the rate at which entanglement can be generated across the weak link \cite{Acoleyen2013,Audenaert2014,Bravyi2007}. This bound states that in a pure state
\be\label{entanglement_rigorous_bound}
\f{\dd S_{Ab_1}}{\dd t}  \leq c || H_{b_1 b_2} || \ln d.
\ee
Here $c$ is a numerical constant given in \cite{Acoleyen2013}; $d$ is the smaller of the Hilbert space dimensions of $b_1$ and $b_2$, here given by $d=2$. Most importantly,  $||H_{b_1b_2}||$ is the magnitude of the largest eigenvalue of $H_{b_1b_2}$, here equal to $J_\text{weak}$.

For small $J_\text{weak}$ the physical rate of entanglement growth (i.e. in a typical state) may be much smaller than the rigorous upper bound (\ref{entanglement_rigorous_bound}). Indeed, numerically we find that in a Floquet Ising spin chain the von Neumann entropy growth rate is of order
\be
\f{\dd S}{\dd t} \sim J_\text{weak}^2 \ln 1/J_\text{weak}
\ee
rather than of order $J_\text{weak}$, see Sec.~\ref{weak_link_sec}.  But for our purposes in this section the upper bound (\ref{entanglement_rigorous_bound}) will suffice.

Microscopically, a Griffiths region does \textit{not} consist of weakly-coupled degrees of freedom, so we cannot immediately apply (\ref{entanglement_rigorous_bound}). A better cartoon is that the Griffiths region consists of `slow' degrees of freedom. To see this, recall that deep in the MBL phase the Hamiltonian may be formulated in terms of `l-bits' --- dressed spin variables whose $Z$-components are strictly conserved~\cite{PhenomenologyOfFullyMBL, SerbynLocalConservationLaws}. This picture is also a useful starting point for considering strongly disordered regions which locally resemble the MBL phase. Since these regions are finite, the dressed spin variables are not strictly conserved, but are instead `slow'.

We can learn how to treat such slow degrees of freedom in the context of a toy model, where the Griffiths region is replaced by a single central spin whose $Z$-component is almost conserved. See Fig.~\ref{weak_link_types}, right. We label the left and right regions by $A$ and $C$ respectively, and the central slow spin by $b$. We denote the Pauli operators for the slow spin by $X$, $Y$, $Z$, and take a Hamiltonian whose $O(1)$ terms all commute with $Z$:
\be
H = H_{Ab}(Z) + H_{bC}(Z) + h_\text{weak} X
\ee
Here the notation $H_{Ab}(Z)$ means that this term acts on the central spin only via its $Z$ operator; it can act arbitrarily on the spins in $A$. For concreteness we take the weak term which breaks conservation of $Z$ to be a transverse field.

It is easy to find an example showing that the \textit{instantaneous} rate of entanglement growth between $A$ and the rest, $\dd S_A/\dd t$,  can be $O(1)$.\footnote{Take $C$ to be empty and $A$ to consist of a single spin, take $H_{bC}$ to be an Ising coupling, and take the initial state to have both spins polarized in the $X$ direction.} However the time-averaged rate is small when $h_\text{weak}$ is small. To see this we relate the physical system to a reference system $Ab_1 b_2 C$ in which the spin $b$ is replaced with two spins. There is a mapping from the Hilbert space of $AbC$ into that of $Ab_1b_2C$ given by
\ba
\ket{ \uparrow}_b& \rightarrow \ket{ \uparrow}_{b_1} \ket{ \uparrow}_{b_2},  &
\ket{ \downarrow}_b& \rightarrow \ket{ \downarrow}_{b_1} \ket{ \downarrow}_{b_2}
\end{align}
in the $Z$ basis. This mapping commutes with the time evolution if we choose the following Hamiltonian for the reference system:
\be
H_\text{reference} = H_{Ab_1}(Z_1) + H_{b_2C}(Z_2) + h_\text{weak} X_1 X_2,
\ee
where $H_{Ab_1}(Z_1)$ is simply $H_{Ab}(Z)$ with $Z$ replaced by $Z_1$.   Further, if two states $\ket{\psi}_\text{phys}$ and $\ket{\psi}_\text{ref}$ are related by the mapping, they yield the same density matrix on $A$. Therefore we are guaranteed that
\be\label{entropy_equality}
S_A^\text{phys} (t) = S_A^\text{ref} (t)
\ee
at all times.

Since the weak field has now become a weak interaction between $Ab_1$ and $b_2 B$, we can use Eq.~\ref{entanglement_rigorous_bound} to bound the change $\Delta S_{Ab_1}$ in the entanglement between $A b_1$ and $b_2 B$:
\be
\Delta S^\text{ref}_{Ab_1}(t)  \leq  (c  \ln 2) \, h_\text{weak} t.
\ee
Of course $S^\text{ref}_{Ab_1}$ is not meaningful in the physical system. But subadditivity of the von Neumann entropy, together with (\ref{entropy_equality}), guarantees that it is close to the quantity of interest: ${| S_{Ab_1}^\text{ref}(t)  - S_{A}^\text{phys} (t)|} \leq \ln 2$. This gives the desired result
\be
\Delta S_A^\text{physical}(t) \leq  ( c \,h_\text{weak}\, t + 2)  \ln 2.
\ee
We see that at long times the time-averaged $\dd S_A / \dd t$ is at most of order $h_\text{weak}$, and that the coefficient remains of $O(1)$ even if the size of $A$ or $C$ diverges. (See Sec.~\ref{weak_link_sec} for a numerical analysis of this problem in a Floquet spin chain, showing that the growth rate is even smaller than the maximum allowed by this bound.)

Finally we turn to a spin chain with subsystems $ABC$, where $B$ is a Griffiths region consisting of a large number $\ell$ of consecutive spins. We will consider a strongly disordered Griffiths region which locally resembles the fully many body localized phase. Write the Hamiltonian as
\be
H = H_A + H_B + H_C + H_{AB} + H_{BC},
\ee
where $H_{AB}$ and $H_{BC}$ each act on a single bond.  Now we make use of the l-bit picture for the MBL phase \cite{PhenomenologyOfFullyMBL, SerbynLocalConservationLaws}. We expect that a unitary transformation on $B$ can reduce its Hamiltonian to a form which only depends on the $Z_i$ operators for spins $i \in B$,
\ba
U H_B U^\dag &=  H'_B,
&
H'_B &= \sum_{i} h_i Z_i + \sum_{ij} J_{ij} Z_i Z_j + \ldots,
\end{align}
where the couplings decay exponentially with distance, and where the unitary transformation preserves the locality of operators up to exponential tails.  For the purposes of entanglement growth we can work with $H'  = U H U^\dag$ instead of $H$, since $U$ only introduces $O(1)$ entanglement. $H'$ is of the form
\be
H' = H_A + H_C + H'_B + U \lf H_{AB} + H_{BC} \ri U^\dag.
\ee
If $H_{AB} = Z_\alpha Z_\beta$, where $\alpha$, $\beta$ are the boundary spins in $A$ and $B$ respectively,
\ba\notag
H_{AB}'  = U H_{AB} U^\dag = Z_\alpha & \bigg(
\sum_{i \in B} (u^X_i X_i + u^Y_i Y_i +u^Z_i Z_i)
\\
&+ \sum_{i,j\in B} k_{ij}^{ZZ} Z_i Z_j + \ldots
\bigg).
\end{align}
Now we split the Griffiths region $B$ into the left-hand, central and right-hand regions, $a$, $b$, and $c$, of length $\ell/3$ each, and apply the duplication trick to $b$ to give a system
\be
A a b_1 b_2 c C.
\ee
We now consider generation of entanglement across the cut $Aab_1 | b_2c C$. The terms in $H_A$ and $H_C$ act only within $A$ or $C$ respectively and can be neglected. Next we have the terms coming from $H_{B}'$. These act on all the $Z$s in $a,b,c$. We have some freedom in how we represent these terms, since  any $Z$ in $b$ can be  represented either by a $Z$ in $b_1$ or by the corresponding $Z$ in $b_2$. Using this freedom, any term in $H_B'$ which does not act on $\textit{both}$  $a$ and $c$ may be represented with a term which does not cross the cut $A a b_1 | b_2 c C$. The largest remaining terms from $H_B'$,  which act on both $a$ and $c$, are of magnitude $\sim \exp{(- \ell / 3\zeta)}$. These terms typically act on $O(\ell)$ spins in region $b$ but only $O(1)$ spins in $a$ and $c$, so can be represented by terms involving only $O(1)$ spins on one side of the cut and $O(\ell)$ on the other. Finally we have the terms from $H_{AB}$ and $H_{BC}$ which couple across the  boundary of the Griffiths region. After the unitary transformation, $H_{AB}$ couples the boundary spin  $\alpha$ in region $A$ to all of the $l$-bits in the Griffiths regions, with exponentially decaying couplings. Terms in which $\alpha$ is coupled to the leftmost spins in region $b$ are of size $\sim \exp{(- \ell/ 3 \zeta)}$. A term like $e^{-\ell /3\zeta} Z_\alpha (...) X_i$,  where $i$ is a site in region $b$, becomes a coupling involving the corresponding sites in both $b_1$ and $b_2$. Again these terms involve $O(\ell)$ spins on one side of the cut and $O(1)$ on the other.

Altogether, the strongest  terms in the Hamiltonian which couple across the cut $A a b_1 | b_2 c C$ have size $\sim e^{-\ell /3\zeta}$ and involve $O(\ell)$ spins.\footnote{These terms can be split into two groups, each of which acts on only $O(1)$ spins from one side. Since (\ref{entanglement_rigorous_bound}) can be applied to each group separately and the results added, the appropriate $\log d$ factor  (\ref{entanglement_rigorous_bound}) is $O(1)$ rather than $O(\ell)$. In any case this polynomial correction is negligible  in comparison to the error in estimating the coefficient in the exponential.}  For a nonrigorous application of the bound (\ref{entanglement_rigorous_bound}), we must estimate the norm of the  entangling Hamiltonian. This may be larger than $e^{-\ell /3\zeta}$ by an exponentially large combinatorial factor, since the number of terms is large. We take the region to be sufficiently strongly disordered (i.e. $\zeta<\zeta_0$, where $\zeta_0$ is an order one constant) that the  exponential decay of the couplings wins, giving
\be
\f{\dd S_{Aab_1}}{ \dd t} \lesssim  e^{-\alpha \ell}
\ee
for some $\alpha$ which does not depend on $\ell$. The difference between $S_{Aab_1}$ and the physically meaningful entropy $S_A$ (or $S_{Aa})$ is $O(\ell)$, and is unimportant at long times.

This shows that the entanglement growth rate across a strongly disordered Griffiths region is exponentially small in the size of the region, regardless of the size of the complete chain. (Assumptions made in previous work are equivalent to taking the entanglement growth rate to grow with the size of the adjacent regions, which we see is not the case.) This is enough to show that there is a power law distribution of local rates. In turn this suffices to prove that $S$ grows subballistically, by the logic in Sec.~\ref{entanglement_growth_sec}. We focussed on  strongly disordered regions  with $\zeta<\zeta_0$, while in practise the prevalent weak links with a given $\Gamma$ may be less strongly disordered. However this will only decrease the value of the exponent $a$.

\section{Entangling across a weak link: numerics}
\label{weak_link_sec}

\begin{figure}[t]%
    \centering
    \subfloat{{\includegraphics[width=0.9\linewidth]{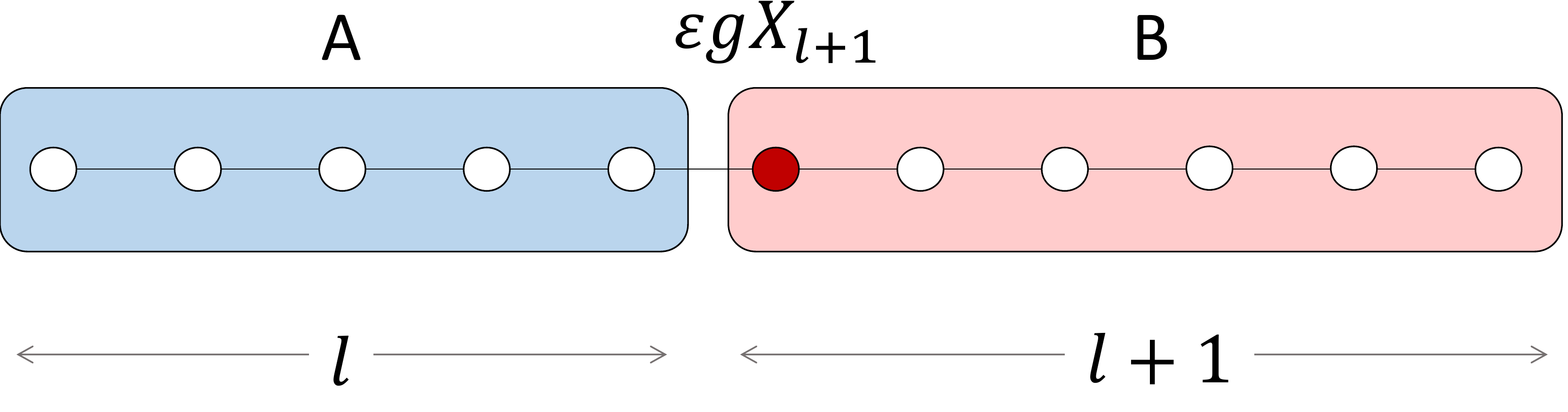} }}%
    \caption{Setup for the numerical simulation. A chain of $L = 2l+1$ spin-$1\over 2$ degrees of freedom is subjected to Floquet dynamics with Ising couplings and transverse and longitudinal fields. The $Z$-component of the central spin is almost conserved due to the weak transverse field.}%
    \label{fig:setup}%
\end{figure}

We now present a numerical study of entanglement growth across a microscopic weak link.
The setup for the numerical simulation is a spin-${1\over 2}$ chain of length $L = 2l+1$, shown schematically  in Fig.~\ref{fig:setup}. We use Floquet dynamics considered previously in Ref. \onlinecite{zhang2015thermalization}: an Ising spin chain with longitudinal and transverse fields. This system has been shown to  thermalize rapidly in the absence of a weak link. A simplifying feature of the Floquet case is that energy is not conserved \cite{zhang2015thermalization}. The time evolution is implemented using the online package ITensor~\cite{ITensor}.

The weak link we consider is formed by the spin at the central site, whose $Z$-component is almost conserved. The commutator of this operator with the one-step time evolution operator is of order $\epsilon$, where $\epsilon$ will be taken to be small. As discussed in Sec.~\ref{explicit_griffiths_argument}, this is a toy model for a Griffiths region: the central spin is an `almost' $l$-bit, whereas all other spins are strongly non-conserved under the evolution. The slow spin bottlenecks the growth of entanglement between the rapidly thermalizing regions surrounding it. We focus on the dynamics of the Renyi entropies (including the von Neumman entropy $S_{\mathrm{vN}} \equiv S_1$)
\be
S_n (t) = {1\over  1-n}\,\log_2\,{\text{Tr} \rho_{A}^n}
\ee
across the bond connecting the spin at site $l$ to the nearly conserved spin ($\rho_A = \mathrm{Tr}_B \rho$ and the regions $A$ and $B$ are defined in Fig.~\ref{fig:setup}).

Here we will give results only for the model with a slow spin, but we have also simulated a spin chain with a weak bond. As might be expected from the mapping between a slow spin and a weak link in Sec.~\ref{explicit_griffiths_argument}, the results are extremely similar~\cite{Nahum_future}.

Our main goal will be to validate two key elements of the picture for entanglement growth discussed above. Firstly, that the entanglement growth rate in the vicinity of a weak link, $x_i$, can be characterized by a {\it local} rate $\Gamma_i$ which is independent of the size of the surrounding regions. Secondly, that in the scaling limit the growth of entanglement entropy is captured by the simple scaling forms discussed in Sec.~\ref{entanglement_growth_sec}. For the present case, the relevant scaling form, if we start from a product state, is:
\be
S(x,t) = \min\big\{  \Gamma t + |x-x_1|
,\,\,
x,\,\, (L-x),\,\, v_E t
 \big\},
\ee
where $x_1$ is the location of the weak link. If we start from a state in which the two subsystems are separately fully entangled,
\be
S(x,t) = \min\big\{  \Gamma t + |x-x_1|
,\,\,
x,\,\, (L-x)
 \big\}.
\ee
In particular, either of these forms implies that for large $L$ and $x$, the entanglement  $S(x_1,t)$ at the location of the weak link has the simple piecewise linear scaling form
\be\label{scaling_form_at_weak_link}
S(x_1,t) =\min \big\{
\Gamma_1 t,\,\, x_1
\big\}.
\ee
Although simple, these  scaling forms are not a priori obvious: they are nontrivial predictions about the universal physics stemming from the results of this paper.

There are many other natural questions about how weakly coupled systems get entangled over time, but we defer these to a future publication \cite{Nahum_future}.

\begin{figure}[t]%
    \centering
    \subfloat{{\includegraphics[width=\linewidth]{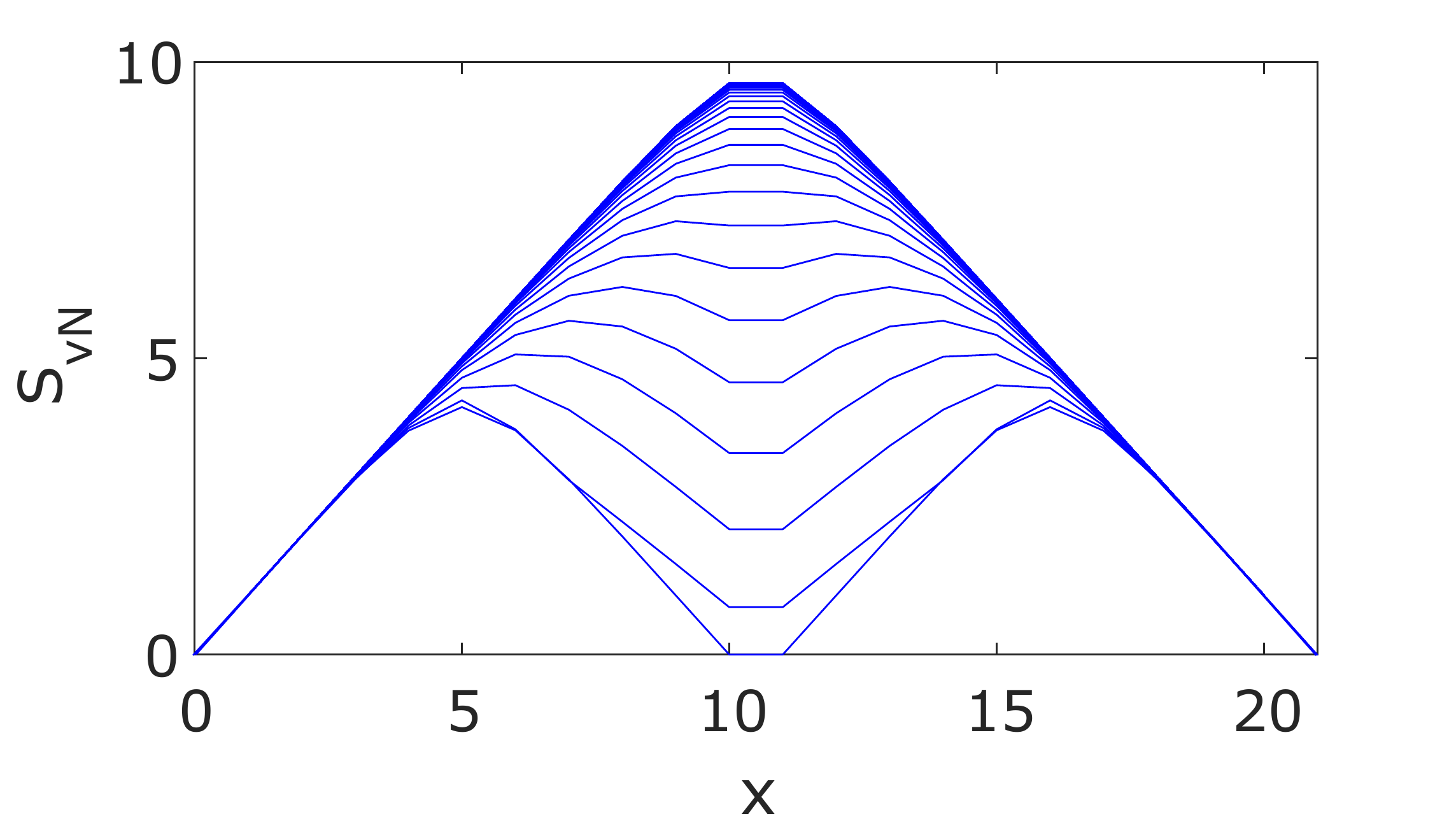} }}%
    \caption{$S_\text{vN}$ as a function of position in systems with $L = 21$ and $\varepsilon = 0.3$. The different curves show times from $t=0$ to $t = 500$ in increments of $\d t = 10$ (the subsystems are separately entangled prior to $t=0$).
    }%
    \label{fig:evolution}%
\end{figure}

\begin{figure}[t]%
    \centering
    \subfloat{{\includegraphics[width=\linewidth]{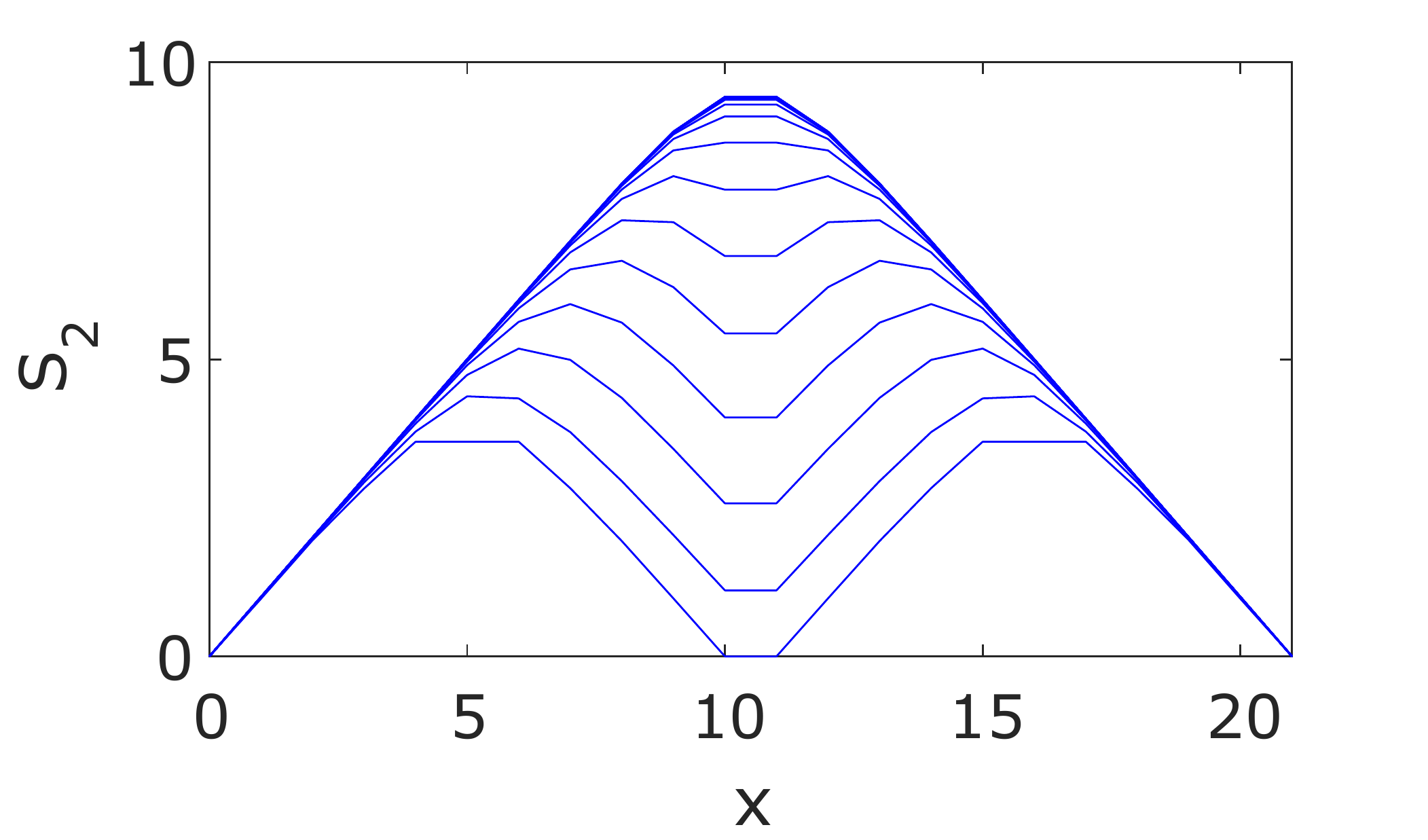} }}%
    \caption{As Fig.~\ref{fig:evolution}, but for the second Renyi entropy $S_2$.
    }%
    \label{fig:evolutionS2}%
\end{figure}

We now give details of the numerical simulation.
We use a Floquet time evolution operator
\be
U(\epsilon) = U_Z U_X(\epsilon),
\ee
which includes half a period of evolution with the Ising interactions and longitudinal fields,
\ba
U_Z = \exp \lf
- i \tau \left[ J \sum_{j=1}^{L-1} Z_j Z_{j+1} + h \sum_{j=1}^L Z_j \right]
\ri,
\end{align}
preceded by half a period of evolution with the transverse fields,
\ba
U_X(\epsilon) = \exp \lf
- i \tau g \left[
\epsilon \, X_{l+1} +
\sum_{j\neq l+1} X_j
\right]
\ri.
\end{align}
Note that the transverse field on the central spin is weakened by a (variable) factor $\epsilon$. The other couplings are fixed at  the values  $J = 1$, $h = 0.809$, $g = 0.9045$, $\tau = 0.8$ which were shown in Ref. \onlinecite{zhang2015thermalization} to yield rapid thermalization.

We start from the product state $\ket{\uparrow \ldots \uparrow}$ in the $Z$ basis. But before beginning the entangling dynamics, we evolve the left and right  subsystems separately, i.e. with $\epsilon=0$, for a time $T_0$ to ensure that they are separately strongly entangled. In the simulations we have used $T_0 = 5$, however, we note that our results depend very weakly on $T_0$. The state at time $t \geq 0$ is therefore defined to be
\be
|\psi(t) \rangle = U(\epsilon) ^{t} \;U(0)^{T_0} \ket{ \uparrow \ldots \uparrow }.
\ee
Note that the entanglement between $A$ and $B$ is zero when $t=0$. In Fig.~\ref{fig:evolution}  we plot a typical spatial dependence of $S_\text{vN}(x,t)$ for a system of size $L=21 $ and for successive times $t$ starting from $t = 0$. Fig.~\ref{fig:evolutionS2} shows the same for $S_2(x,t)$.

\begin{figure}[h]%
    \centering
    \subfloat{{\includegraphics[width=\linewidth]{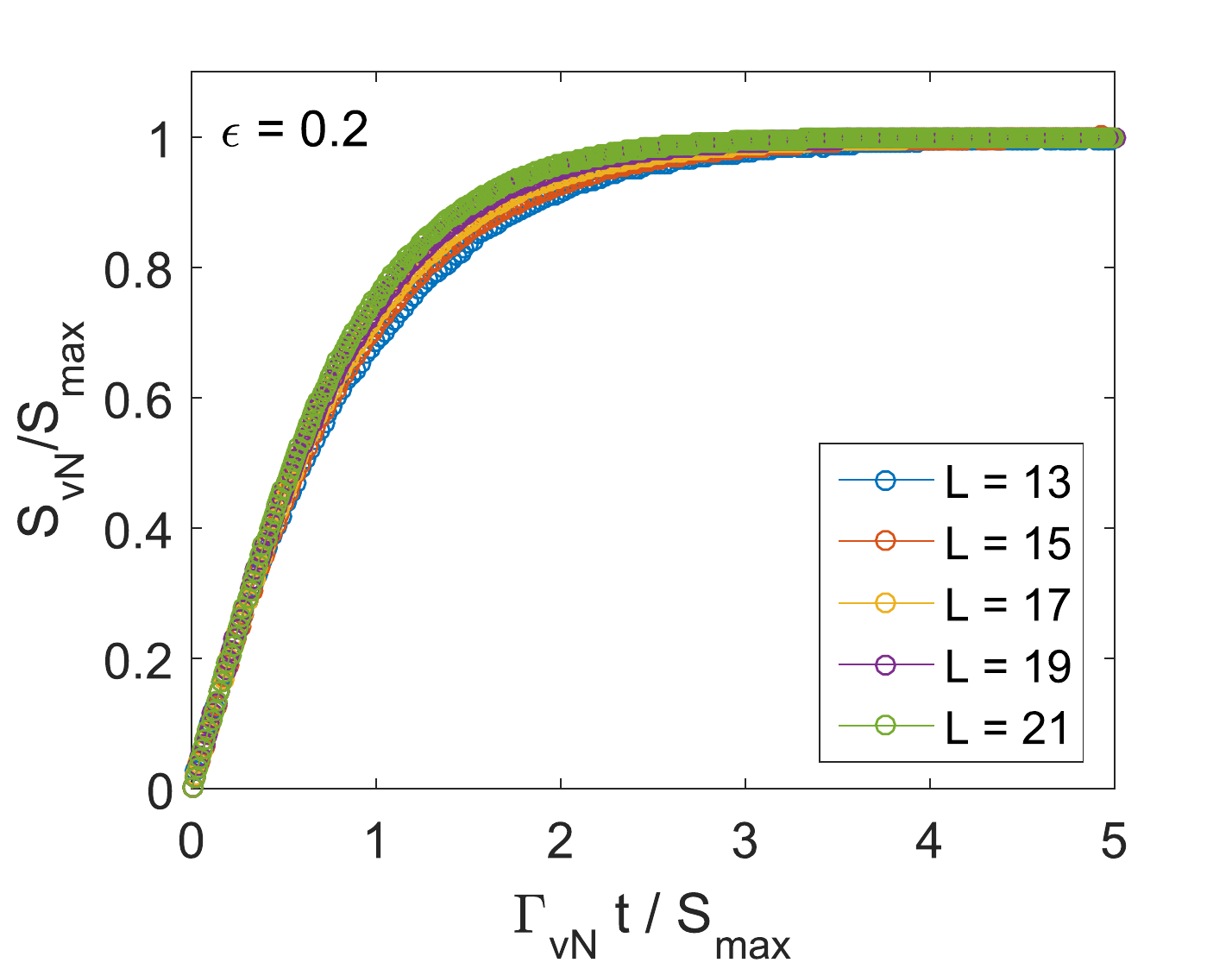} }}%
    \qquad
    \subfloat{{\includegraphics[width=\linewidth]{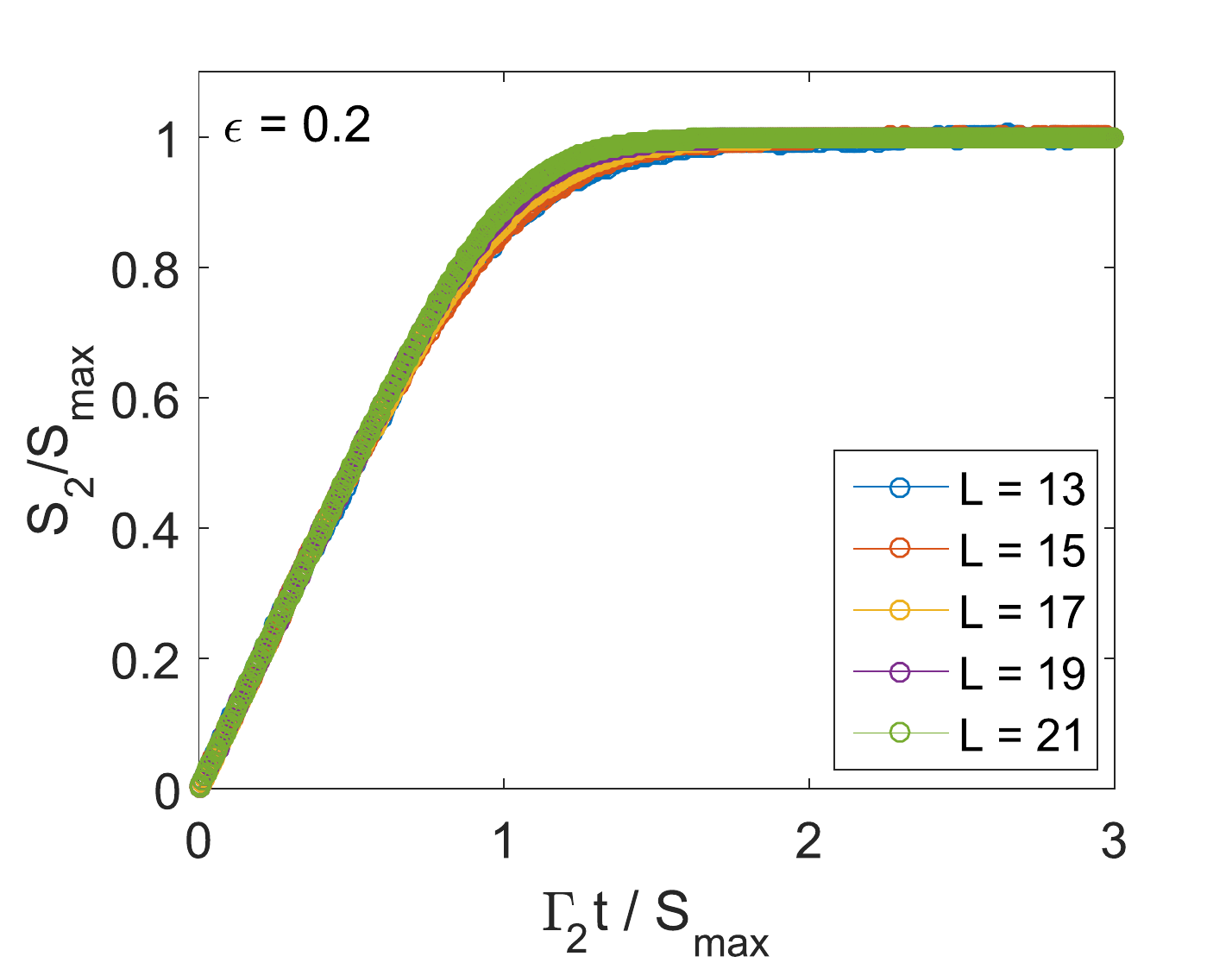} }}%
    \caption{The von Neumann (top) and second Renyi (bottom) entropies scaled by their maximal value at large time $S_{max}$ vs. the rescaled time $\Gamma_n t / S_{max}$ for different values of $L$ and for $\epsilon = 0.2$. The value of $S_{max}$ is determined numerically.}%
    \label{fig:scaling}%
\end{figure}

Let us now examine the validity of the scaling forms. In Fig.~\ref{fig:scaling} we plot the the von Neumann and second Renyi entropies, rescaled by their asymptotic late time values, against the rescaled time variable $\Gamma_n t / S_\text{max}$. Here $\Gamma_n$ is the growth rate of $S_{n}(t)$ obtained by fitting the linear growth near $t = 0$ for $L = 21$. $S_\text{max}$ is the value of the entropy as $t \rightarrow \infty$, which is determined numerically and which depends on $n$ and $L$.  We note that in the case of the von Neumann entropy $S_\text{max}$ converges very quickly with increasing $L$ to the prediction of Ref. \onlinecite{Page1993} for a random state.

\begin{figure}[t]%
    \centering
    \subfloat{{\includegraphics[width=0.9\linewidth]{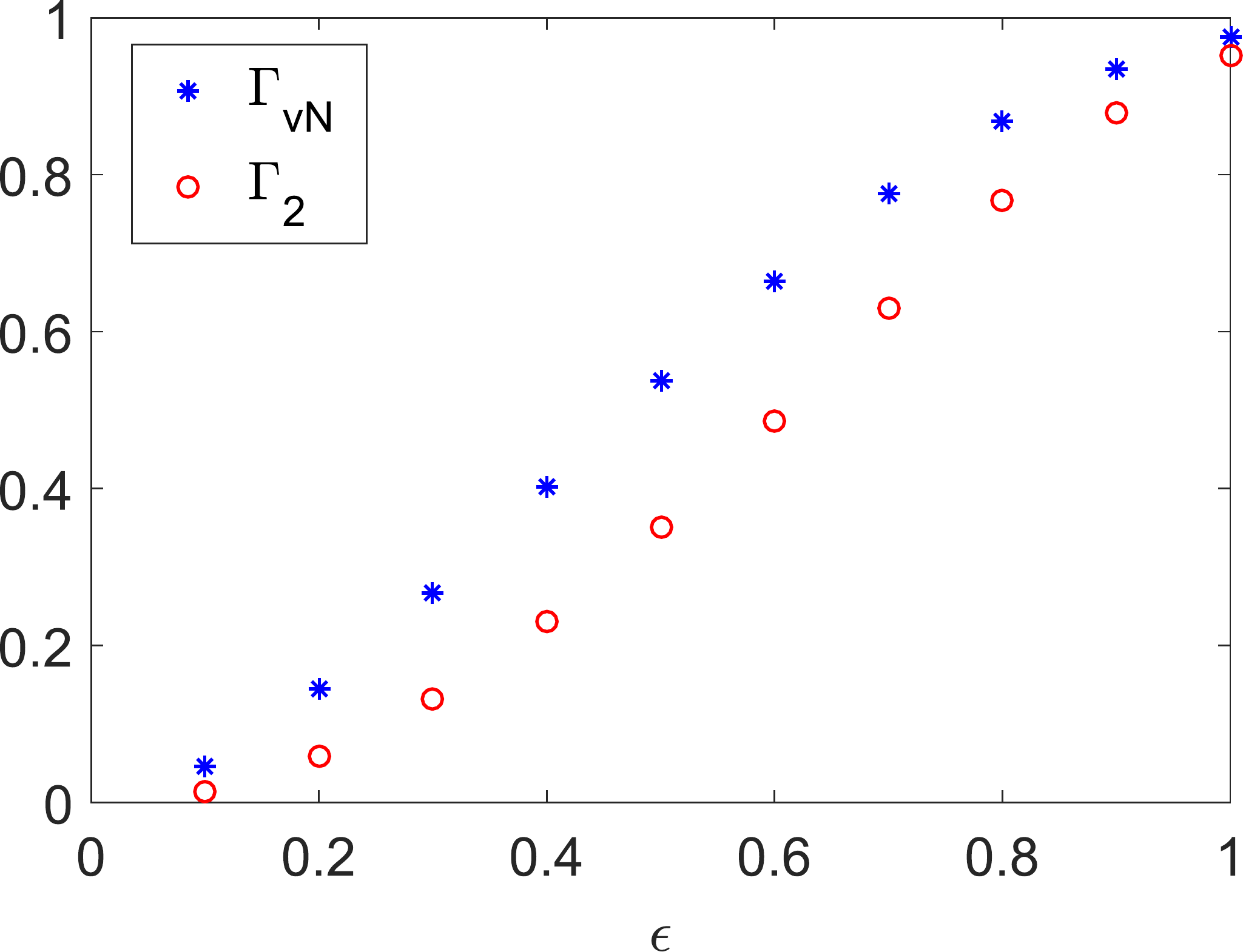} }}%
    \caption{The initial growth rates $\Gamma_n$ as a function of $\epsilon$ (calculated using $L = 21$). The stars (blue) and circles (red) correspond to von Neumann and second Renyi entropy, respectively.}%
    \label{fig:initial_rate}%
\end{figure}

It is clear from Fig.~\ref{fig:scaling} that the early time growth rate, $\Gamma_n$, is  independent of the length of the system. This is consistent with our coarse-grained picture. By numerically fitting the data we find that for small $\epsilon<0.1$ the growth rates are given by
\ba
 \Gamma_{\mathrm{vN}} & = A_{\mathrm{vN}} \, \epsilon^2 \, \log {{B} \over \epsilon},
&
\Gamma_n  &= \f{A \, n}{n-1} \epsilon^2,
 \label{linear growth rate}\end{align}
where $A_{\mathrm{vN}},\,B$ and $A$ are $\epsilon$--independent constants.
The rates for larger values of $\epsilon = 0.1,0.2,\ldots,1$ are shown in Fig.~\ref{fig:initial_rate}.
This confirms that indeed the growth rate is small in $\epsilon$ and that a nearly conserved quantity causes a bottleneck for entanglement growth. Note that for small $\epsilon$ the growth rate is  parametrically smaller than the rigorous upper bound in Sec.~\ref{explicit_griffiths_argument}.  Interestingly, Fig.~\ref{fig:initial_rate} suggests that for $\epsilon=1$ (the homogeneous case) the growth rates for $S_\mathrm{vN}$ and $S_2$ may be equal.

We also find that the transition from linear growth to saturation becomes sharper with increasing $L$, consistent with the expected scaling form Eq.~(\ref{scaling_form_at_weak_link}). This is much clearer for the second and higher Renyi entropies (Fig.~\ref{fig:scaling}). For the von Neumann entropy a more careful finite size analysis is required, which we give below.

\begin{figure}[t]%
    \centering
    \subfloat{{\includegraphics[width=0.505\linewidth]{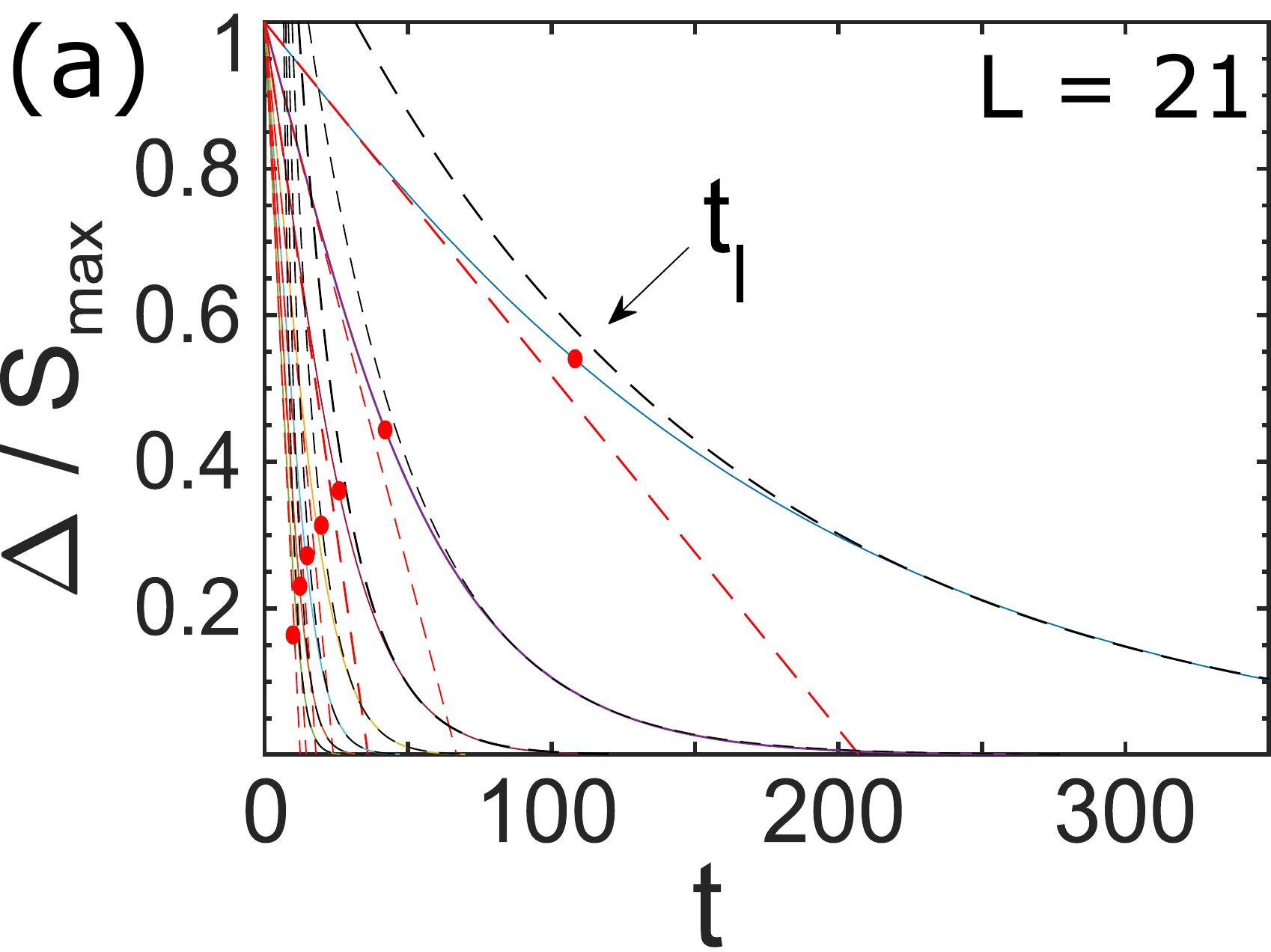} }}%
    \subfloat{{\includegraphics[width=0.495\linewidth]{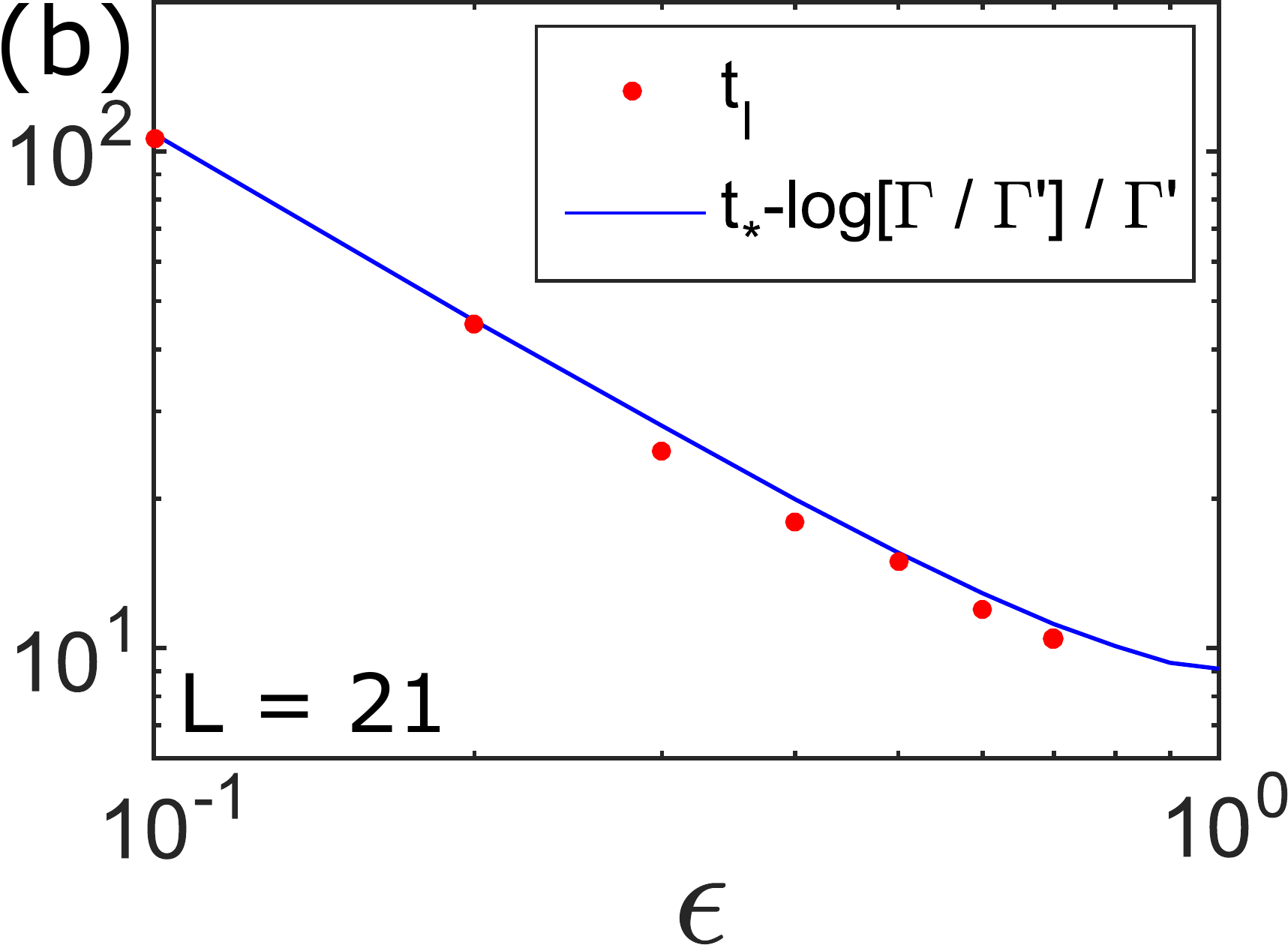} }}%
    \caption{ (a): The {\it solid} lines show the normalized deviation from maximal entropy $\Delta / S_\text{max}$ defined in Eq.~(\ref{exponential growth}) vs. time, $t$. The different colors correspond to different values of $\epsilon$ ranging between $0.1,0.2,\ldots,0.7$.  The {\it dashed} curves are the fits to the linear form: $1-\Gamma t /S_{\text{max}} $ (red dashed) and to the exponential form: $1-\exp\left[ -\Gamma'\left(t-t_* \right)\right]/S_{\text{max}}$ (black dashed). The red dots mark the {\it crossover points} between linear behavior, at short times, and exponential, at long times. These crossover points are used to extract the time scale $t_l$ is extracted. (b): Comparison between the extracted crossover time $t_l$ (red dots) and the expected form Eq.~(\ref{t_l}) (blue solid curve). }%
    \label{fig:tl}%
\end{figure}

For translationally-invariant Floquet systems with no conserved quantities, the late time saturation of the entropy has been found to be exponential \cite{zhang2015thermalization}. We also obtain a good fit at late times with
\ba \label{exponential growth}
S_{\mathrm{vN}} &= S_\text{max} -  \Delta(t), &
\Delta(t) &= \exp\left(-\Gamma' | t-t_*|  \right),
\end{align}
where the parameters $\Gamma'$ and $t_*$ of the fit are extracted separately for each $L$ and $\epsilon$. We find
 that at small $\epsilon$ the saturation rate $\Gamma'$ behaves as
\be
\Gamma' = A' \epsilon^2. \label{exponential growth rate}
\ee
The time scale $t_*$  for saturation is of order $L$.

Assume that there is a simple crossover between linear and exponential behaviour at a time $t_l$. Matching $S$ and $\partial S/\partial t$ at this time gives the following gluing conditions:
\ba
t_l & = {S_\text{max} \over \Gamma_{\mathrm{vN}}} - {1\over \Gamma'}, &
t_* & = t_l +{1\over \Gamma'} \log {\Gamma_{\mathrm{vN}}\over \Gamma' } \label{t_l},
\end{align}
which means that
\be S(t_l) = S_\text{max} - \Gamma_{\mathrm{vN}} / \Gamma'.
\ee
In Fig. \ref{fig:tl} we test the  assumption underlying the previous. In the first panel we present the linear (red dashed lines) and exponential (black dashed lines) fits to $\Delta(t)/S_\text{max}$. The red dots are crossover times between these two behaviors, determined by eye. In the second panel we compare these crossover times with the prediction in Eq.~(\ref{t_l}), finding good agreement.

The previous equation shows that the departure from linear behaviour occurs when the entropy is within
\be
\Delta S_\text{vN} \sim \f{\Gamma_{\mathrm{vN}}}{ \Gamma'}
\sim \f{A_{vN}}{A'} \log \f{B}{\epsilon}.
\ee
bits of its maximum. The $\Delta S \sim \log 1/\Gamma$ behaviour for the von Neumann entropy indicates that the finite-size rounding is more severe for weaker weak links; this conforms with what we find numerically. For the higher Renyi entropies this logarithmic factor is absent, as a result of Eq.~\ref{linear growth rate}. Since $\Delta S$ is independent of $L$, the above confirms the scaling form Eq.~\ref{scaling_form_at_weak_link} for a system of size $L$ with a weak link whose strength is fixed as $L\rightarrow \infty$.  (The deviation from linearity for the last $\Delta S \sim  \log 1/\Gamma$ bits of entanglement is also negligible if we take $\Gamma$ to scale as a negative power of $L$, which is the natural scaling in the Griffiths regime.) 

Note that if the relative logarithmic factor in the growth rates for $S_\text{vN}$ and $S_{n>1}$ (Eq.~\ref{linear growth rate}) also applies for coarse-grained weak links, then  the von Neumman entropy in the Griffiths phase will exceed the higher Renyi entropies by a factor of order $\log t$ at late times (since the weak links that are relevant at time $t$ have strength  ${\Gamma \sim t^{-1/(2+a)}}$).

We have confirmed the two key assumptions: $\Gamma_n$ is independent of $L$, and the scaling form Eq.~(\ref{coarse-grained growth rule}) is indeed obtained in the large $L$ limit for a microscopic weak link. If we allow ourselves to extrapolate the above results to Griffiths regions (coarse-grained weak links), these conclusions  support the scaling forms derived from the surface growth picture.

\section{Summary}
\label{conclusions_sec}

{
We have examined the dynamics of isolated one-dimensional quantum systems that thermalize, but  have a distribution of weak links produced by quenched randomness.  This occurs in systems which sustain (but which are outside) the many-body-localized (MBL) phase, due to Griffiths regions that are locally in this phase.  We developed coarse-grained pictures for the production of entanglement and the spreading of operators and of conserved densities.  

We found that such systems have multiple dynamic length scales that diverge with different powers of the time.  These length scales are for: the growth of entanglement, the spreading of conserved densities, the spreading of initially local operators, and the width of the `front' of a spreading operator.  Contrary to  systems without quenched disorder, the process of entanglement saturation (either for states or operators) is always parametrically slower than the spreading of operators as measured by the norm of their commutator with a local operator or by the out-of-time-order correlator.

We also examined the exchange of quantum information between two clean systems coupled by a single weak link, giving universal scaling forms for the growth of entanglement and the spreading of operators from one system into the other. 

Various questions remain for the future. It has been suggested on the basis of simulations \cite{barlevquasiperiodic} that sub-ballistic entanglement growth may occur in the thermal phase of 1D chains with quasiperiodic detuning rather than disorder. The relation of such systems to rare-region physics remains to be understood.

The pictures described here are set up to deal with dynamics in the thermal phase. It would be interesting to understand the crossover effects at the MBL phase transition \cite{VoskHuseAltmanMBL,PotterVasseurParameswaranMBL,KhemaniCriticalProperties2017}  in  more detail. It may also be interesting to consider a coarse-grained treatment of the network of locally \textit{thermal} Griffiths regions in the MBL phase. In higher dimensions, such rare regions have been argued to destabilize the MBL phase \cite{de2017stability,de2017many}.
}

\begin{acknowledgments}
We thank S. Gopalakrishnan, J. Haah, C. Jonay,  and S. Vijay for collaboration on related work.  AN acknowledges support from the Gordon and Betty Moore Foundation under the EPiQS initiative (grant No.~GBMF4303) and from EPSRC Grant No.~EP/N028678/1. JR acknowledges a fellowship from the Gordon and Betty Moore Foundation under the EPiQs initiative (grant No.~GBMF4303).
\end{acknowledgments}

\bibliography{entanglement-ref-Griffiths}

\end{document}